\documentclass[aps,twocolumn,showpacs,floatfix,eqsecnum,prb]{revtex4-1}

\usepackage{amsmath}
\usepackage{graphicx}% Include figure files

\begin{document}

\title{From Luttinger liquid to non-Abelian quantum Hall states}

\author{Jeffrey C.Y. Teo}

\altaffiliation{Present Address: Department of Physics, University of Illinois at Urbana-Champaign,
1110 West Green Street, Urbana, Illinois 61801-3080.}

\affiliation{Department of Physics and Astronomy, University of Pennsylvania, Philadelphia, PA 19104}

\author{C.L. Kane}
\affiliation{Department of Physics and Astronomy, University of Pennsylvania, Philadelphia, PA 19104}

%\affiliation{Department of Physics and Astronomy, University of Pennsylvania, Philadelphia, PA 19104}

\begin{abstract}
We formulate a theory of non-Abelian fractional quantum Hall states by considering an anisotropic system consisting of coupled, interacting one dimensional wires.  We show that Abelian bosonization provides a simple framework for characterizing the Moore Read state, as well as the more general Read Rezayi sequence.  This coupled wire construction provides a solvable Hamiltonian formulated in terms of electronic degrees of freedom, and provides a direct route to characterizing the quasiparticles and edge states in terms of conformal field theory.   This construction leads to a simple interpretation of the coset construction of conformal field theory, which is a powerful method for describing non Abelian states.  In the present context, the coset construction arises when the original chiral modes are fractionalized into coset sectors, and the different sectors acquire energy gaps due to coupling in ``different directions".  The coupled wire construction can also can be used to describe anisotropic lattice systems, and provides a starting point for models of fractional and non-Abelian Chern insulators.
This paper also includes an extended introduction to the coupled wire construction for Abelian quantum Hall states, which was introduced earlier.

\end{abstract}

\maketitle

\section{Introduction}
\label{sec:I}

The search for non-Abelian states in electronic materials is an exciting frontier in condensed matter physics\cite{qcreview}.  Motivation for this search is provided by Kitaev's proposal\cite{kitaev} to use such states for topological quantum computation.  The quantum Hall effect is a promising venue for non Abelian states.  There is growing evidence that the pfaffian state introduced by Moore and Read\cite{mr,readgreen} describes the quantum Hall plateau observed at filling $\nu = 5/2$\cite{greiter,heiblum,marcus,willett}.  The Moore Read state gives the simplest non-Abelian state, with quasiparticles that exhibit Ising non-Abelian statistics.  While the observation and manipulation of Ising anyons is an important goal, Ising anyons are not sufficient for universal quantum computation\cite{freedman}.    The $Z_3$ parafermion state introduced by Read and Rezayi\cite{readrezayi} is a candidate for the quantum Hall plateau at $\nu=12/5$.  The quasiparticles of the Read Rezayi state are related to Fibonacci anyons\cite{slingerland,trebst}, which have a more intricate structure that in principle allows universal quantum computation\cite{freedman,bonesteel}.

There is currently great interest in realizing quantum Hall physics in materials without an external magnetic field or Landau levels.  This possibility was inspired by Haldane's realization\cite{haldane} that a zero field integer quantum Hall effect can occur in graphene, provided time reversal symmetry is broken.  Though such an anomalous quantum Hall effect has not yet been observed, related physics occurs in topological insulators\cite{hasankane,qizhang}, which have been predicted and observed in both two and three dimensional systems.  Recently, there have been suggestions for generalizations of this idea to zero field fractional quantum Hall states\cite{neupert,sheng,qiqhall,regnault}, as well as fractional topological insulators\cite{levinstern,maciejko,swingle,levin2}.
%Topological superconductivity, implemented by a $p+ip$ superconductor or proximity effect %devices, provides a promising route to a zero field non Abelian state.
%These systems are predicted to exhibit Majorana fermion states which can in principle be %manipulated and measured.  However, there is a fundamental difference between topological %superconductors and the Moore Read state.  In the Moore Read state, the quasiparticles that %exhibit Ising non Abelian statistics are quantum particles that can interfere.  In %contrast, the non-Abelian statistics in topological superconductors is associated with %classical vortex configurations which can not interfere.
The question naturally arises whether it is possible to engineer zero field non-Abelian quantum Hall states\cite{kapit,bernevig2}, which exhibit the full Ising, or even Fibonacci anyons.

A difficulty with answering this question is a lack of methods for dealing with strongly interacting systems.  Moore and Read, and later Read and Rezayi, built on Laughlin's idea\cite{laughlin} and constructed trial many body wave functions as correlators of non trivial conformal field theories\cite{mr,readrezayi}.  This allowed most properties of the quasiparticles and edge states to be deduced, and it had the virtue of allowing the construction of interacting electron Hamiltonians with the desired ground state.
However, since this approach relied on the structure of the lowest Landau level, it is not clear how it can be applied to a lattice system.
Effective topological quantum field theories\cite{girvin, zhang,read,lopez,wenzee} and parton constructions
\cite{wenparton} provide an elegant framework for classifying quantum Hall states and provide a description of their low energy properties.  However, since the original electronic degrees of freedom are replaced by more abstract variables, these theories provide little guidance for what kind of electronic Hamiltonian can lead to a given state.

In this paper we introduce a new method for describing non Abelian quantum Hall states by considering an anisotropic system consisting of an array of coupled one dimensional wires.  Study of the anisotropic limit of quantum Hall states dates back to Thouless et al.\cite{tknn}, who used this limit to evaluate the Chern invariant in the integer quantum Hall effect.  The Chalker Coddington model\cite{chalker} for the integer quantum Hall effect also has a simple anisotropic limit, which is closely related to the coupled wire model.
A coupled wire construction for Abelian fractional quantum Hall states was introduced in Ref. \onlinecite{kml}.  
Here we build on that work and show how the coupled wire construction can be adapted to describe the Moore Read pfaffian state, as well as the more general Read Rezayi sequence of quantum Hall states.

The coupled wire construction has a number of desirable features.   First, it allows for the definition of a simple Hamiltonian, expressed in terms of electronic degrees of freedom, that can be transformed, via Abelian bosonization, into a form that for certain special
parameters can be solved essentially exactly.  The quasiparticle spectrum, as well as the edge state structure follow in a straightforward manner.  The coupled wire construction thus provides a direct link between a microscopic electron Hamiltonian and the low energy conformal field theory description of the edge states.  As such, it provides an intermediate between the wave function approach to quantum Hall states and the effective field theory approach.

The coupled wire construction also provides a simple picture for the quantum entanglement present in quantum Hall states\cite{dong,lihaldane,qikatsura}.  When the electron Hamiltonian is transformed via Abelian bosonization, it becomes identical to a theory of strips of quantum Hall fluid coupled via electron tunneling between their edge states.  It thus provides a concrete setting for the more abstract coupled edge state models considered by Gils, et al.\cite{gils}.  The coupled wire model is also similar in spirit to the AKLT model of quantum spin chains\cite{aklt}, which provides a similarly intuitive and solvable model for understanding fractionalization in one dimension.  For the non Abelian quantum Hall states, the coupled wire construction provides a concrete interpretation for the coset construction, which is a powerful (albeit abstract) mathematical tool for describing non Abelian quantum Hall states\cite{fradkin}.

A final virtue of the coupled wire construction is that it can be applied to zero field anisotropic lattice models.  The effect of the magnetic field in the coupled wire model is to modify the momentum conservation relations when electrons tunnel between wires.  A similar effect could arise due to scattering from a periodic potential.  The coupled wire construction may thus provide some guidance for the construction of lattice models for the fractional and non Abelian quantum Hall states.  Note that our construction is somewhat different from the proposals for Chern insulators Refs. \onlinecite{neupert,sheng,qiqhall,regnault} because we do not require nearly flat bands with an non zero Chern number.

The outline of the paper is as follows.  We will begin in section II with an extended introduction to the coupled wire construction for Abelian quantum Hall states.  Much of this material was contained either explicitly or implicitly in Ref. \onlinecite{kml}.  Here, since we are free from the constraints of a short paper, we will fill in some details that were absent in Ref. \onlinecite{kml}.  In particular, we will explain the generalization of the coupled wire construction to describe systems of bosons, we will demonstrate the Abelian fractional statistics of quasiparticles described in our approach, and we will explicitly construct second level hierarchical fractional quantum Hall states.

Section III is devoted to the Moore Read state.  We will begin with a construction of this state for bosons at filling $\nu= 1$. This leads to a bosonized model that can be solved using via fermionization.  We will then show that for a special set of parameters the model has a particularly simple form, which can be interpreted in the framework of the coset construction of conformal field theory.  We will conclude section III by showing how to construct the more general Moore Read state at filling $\nu = 1/(1+q)$ for $q$ even (odd) for bosons (fermions).

In section IV we will generalize our construction to describe the Read Rezayi sequence at level $k$.  Again, the formulation is simplest for bosons at filling $\nu = k/2$, where the coupled wire model is closely related to the coset construction of these state.  We show that the coupled wire model leads maps to a bosonized representation of the critical point of a $Z_k$ statistical mechanics model, which for $k=3$ reduces to the 3 state Potts model.  This bosonized representation allows us to identify the $Z_k$ parafermion primary fields and fully characterize the edge states of the Read Rezayi states.  Finally, as in section II, we conclude by generalizing our results to describe bosonic (fermionic) level $k$ Read Rezayi states at filling $k/(2+ q k)$ for $q$ even (odd).

Some of the technical details are presented in the appendices.  Appendix A gives a careful treatment of Klein factors, while Appendices B and C contain some of the conformal field theory calculations discussed in section IV.

\section{Abelian Quantum Hall States}
\label{sec:II}

\subsection{Coupled Wire Construction for Fermions}

In this section we review the coupled wire construction for fermions introduced in Ref. \onlinecite{kml}. We begin by considering an array of identical uncoupled spinless non interacting one dimensional wires, as shown in Fig. 1a, with a single particle electronic dispersion $E(k)$.  We assume each wire is filled to the same density, characterized by Fermi momentum $k_F$.  The two dimensional electron density is then $n_e = k_F/\pi a$, where $a$ is the separation between wires.  A perpendicular magnetic field, represented in the Landau gauge ${\bf A} = -B y \hat x$, shifts the momentum of each wire.  The right and left moving Fermi momenta of the $j$'th wire are then
\begin{equation}
k^{R/L}_{F j} = \pm k_F^0 + b j,
\end{equation}
where $b = |e|aB/\hbar$.   The filling factor $h n_e/|e| B$ is then
\begin{equation}
\nu=2k_F/b.
\label{nu}
\end{equation}
The low energy Hamiltonian, linearized about the Fermi momenta is
\begin{eqnarray}
%{\cal H}_0 = \sum_{j,a=R/L=\pm} \int dx a v_F^0 \psi^{a\dagger}_j (-i\partial_x - k_{Fj}^a)\psi^a_j,
{\cal H}_0 = \sum_j \int dx v_F^0\left( \psi^{R\dagger}_j (-i\partial_x - k_{Fj}^R)\psi^R_j \right.\nonumber  \\  -\left.
\psi^{RL\dagger}_j (-i\partial_x - k_{Fj}^L)\psi^L_j\right),
\end{eqnarray}
where $\psi^{R/L}_j$ describe the fermion modes of the $j$'th wire in the vicinity of the Fermi points
$k_{Fj}^{R/L}$.

We next bosonize by introducing bosonic fields $\varphi_j(x)$ and $\theta_j(x)$ that satisfy
\begin{equation}
[\partial_x\theta_j(x),\varphi_{j'}(x')] = i \pi  \delta_{jj'}\delta_{xx'},
\label{thetaphicomm}
\end{equation}
where we use the shorthand notation $\delta_{xx'}=\delta(x-x')$.
$\varphi_j(x)$ is a bosonic phase field, while $\theta(x)$ describes density fluctuations.  The long wavelength density fluctuations on the $j$'th wire are
\begin{equation}
\rho_j(x) = \sum_a \psi_j^{a\dagger}(x) \psi^a_j(x) = \partial_x\theta_j(x)/\pi.
\end{equation}
The electron creation and annihilation operators may be written
\begin{eqnarray}
\psi^R_j(x) &=& {\kappa_j\over\sqrt{2\pi x_c}} e^{i\left(k^{R}_{F j} x + \varphi_j(x) + \theta_j(x)\right) }, \\
\psi^L_j(x) &=& {\kappa_j\over\sqrt{2\pi x_c}} e^{i\left(k^{L}_{F j} x + \varphi_j(x) - \theta_j(x)\right) }.  \nonumber
\label{bosonize psirl}
\end{eqnarray}
where
$x_c$ is a regularization dependent short distance cutoff and $\kappa_j$ is a Klein factor that assures the anticommutation of the fermion operators on different wires.  Eq. \ref{thetaphicomm} hides the zero momentum parts of $\theta_j$ and $\phi_j$, which must be accounted for in order to
correctly treat the Klein factors.  Since this issue tends to obscure the simplicity of our construction, we will not dwell on it in the text of the paper.  Appendix A contains a careful discussion of the zero modes and Klein factors, which shows when they can be safely ignored.

In terms of the density and phase variables, the Hamiltonian for non interacting electrons is
\begin{equation}
{\cal H}^0_{\rm non \ int.} = {v_F\over{2\pi}} \sum_j \int dx \left[ (\partial_x\varphi_j)^2 + (\partial_x\theta_j)^2 \right].
\label{h0 free fermion}
\end{equation}

Interactions between electrons as well as electron tunneling between the wires can be added. In general, there are two classes of terms : forward scattering and inter channel scattering.  The forward scattering terms conserve the number of electrons in each channel and can be expressed as the interactions between densities and currents.  This leads to a Hamiltonian that is quadratic in the boson variables,
\begin{equation}
{\cal H}^0_{\rm SLL}[\theta,\varphi] = \sum_{jk} \int dx
\left(\begin{array}{cc}
    \partial_x\varphi_j & \partial_x\theta_j
\end{array} \right)
{\bf M}_{jk}
\left(\begin{array}{c}
        \partial_x\varphi_k \\ \partial_x\theta_k
\end{array}\right).
\label{h0sll}
\end{equation}
Here the $2\times 2$ matrix ${\bf M}_{jk} = \delta_{jk} {\bf I} v_F/2\pi + {\bf U}_{jk}$, where ${\bf U}_{jk}$ describes the forward scattering interactions.  ${\cal H}^0_{\rm SLL}$ describes a gapless anisotropic conductor in a sliding Luttinger liquid phase\cite{ohern,emery,vishwanath,sondhi,ranjan}.

\begin{figure}
\includegraphics[width=3in]{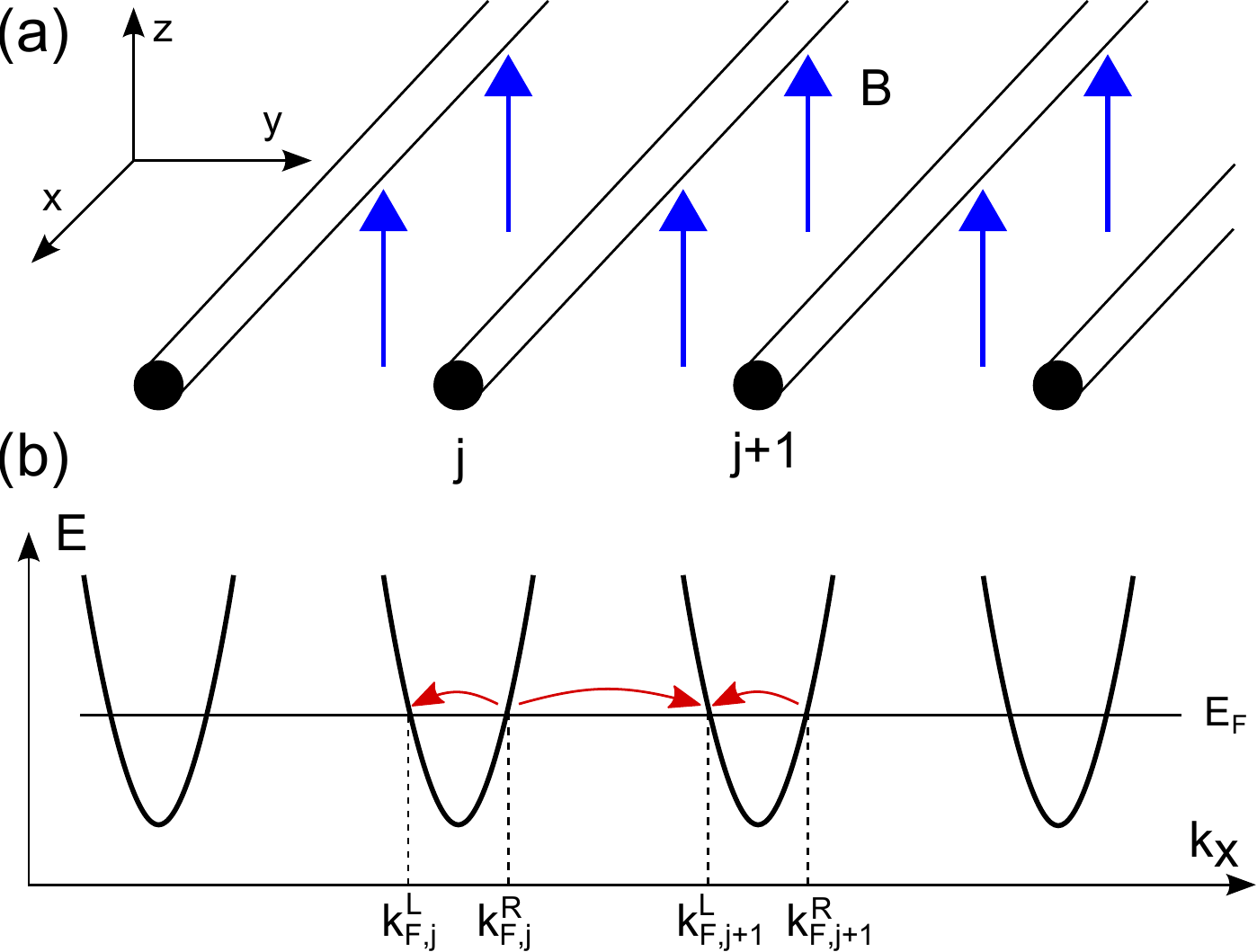}
\caption{(a) An array of coupled wires in a perpendicular magnetic field.  (b)  The magnetic field
shifts the momentum of the wires.  At at special filling factors, there exist momentum conserving correlated tunneling processes that lead to quantum Hall states.  The process shown describes the
Laughlin state at filling $\nu=1/3$.}
\label{figure1}
\end{figure}

Symmetry allowed inter channel scattering terms must be added to ${\cal H}^0_{\rm SLL}$.  They can open a gap and lead to interesting phases.  The allowed terms are built from products of single electron operators, and have the form
\begin{equation}
{\cal O}^{\{m_p,n_p\}}_j(x) =  \prod_p \psi^R_{j+p}(x)^{s_p^R} \psi^L_{j+p}(x)^{s_p^L},
\end{equation}
where $s_p^{R/L}$ are integers such that $\psi^{R/L}_{j+p}$ ($\psi^{R/L\dagger}_{j+p})$ appears $|s_p^{R/L}|$ times for $s_p^{R/L} > 0 \ (<0)$.   It is convenient to write $s_p^{R/L}$ in terms of a new set of integers,
\begin{eqnarray}
s_p^R &=& (n_p + m_p)/2 \\
s_p^L &=& (n_p - m_p)/2.
\end{eqnarray}
Then ${\cal O}^{\{m_p,n_p\}}_j(x)$ takes the form
\begin{equation}
%{\cal O}^{\{m_p,n_p\}}_j(x) =  K^{\{m_p,n_p\}}_j(x)e^{i\sum_p m_p \varphi_{j+p}(x)+n_p\theta_{j+p}(x)}.
%{\cal O}_{j,\{M_p,N_p\}}(x) =  \kappa_{j,\{M_p\}}e^{i \sum_p M_p(bpx + \varphi_{j+p}(x))+N_p(k_Fx+\theta_{j+p}(x))}.
{\cal O}^{\{m_p,n_p\}}_j = c \tilde\kappa_j R \exp i\left[\sum_p n_p \varphi_{j+p}+m_p\theta_{j+p}\right].
\end{equation}
where $c$ is a non universal constant.  The product of Klein factors is
\begin{equation}
\tilde\kappa^{\{m_p,n_p\}}_j =  \prod_p \kappa_{j+p}^{n_p}.
\end{equation}
The oscillatory factor describing the net momentum of ${\cal O}^{\{m_p,n_p\}}_j$ is
\begin{equation}
R^{\{m_p,n_p\}}(x) = \exp i\left(\sum_p  b p n_p + k^0_F m_p\right)x .
\label{R(x)}
\end{equation}
The operators ${\cal O}^{\{m_p,n_p\}}_j$ define a term in the Hamiltonian,
\begin{equation}
V^{\{m_p,n_p\}} = \sum_j \int dx \left(v^{\{m_p,n_p\}} {\cal O}^{\{m_p,n_p\}}_j(x) + h.c.\right).
\end{equation}

There are physical constraints on the allowed $\{m_p,n_p\}$.  Since $s_p^{R/L}$ must be integers, we require
\begin{equation}
m_p= n_p  \ {\rm mod}\ 2.
\label{fermionmn}
\end{equation}
Charge conservation requires that
\begin{equation}
\sum_p n_p = 0.
\label{charge conservation}
\end{equation}
Momentum conservation implies
\begin{equation}
\sum_p (bp n_p + k_F m_p)  = 0,
\label{momentum conservation}
\end{equation}
so that the oscillatory term in \eqref{R(x)} vanishes.

The Hamiltonian
\begin{equation}
{\cal H} = {\cal H}^0_{\rm SLL} + \sum_{\{m_p,n_p\}} V^{\{m_p,n_p\}}
\label{h0sll+v}
\end{equation}
can be studied perturbatively using the standard renormalization group analysis.
The lowest order RG flow equation for $v^{\{m_p,n_p\}}$ is
\begin{equation}
dv^{\{m_p,n_p\}}/d\ell = (2-\Delta^{\{m_p,n_p\}})v^{\{m_p,n_p\}}.
\label{dvdl}
\end{equation}
The scaling dimension $\Delta^{\{m_p,n_p\}}$ depends on the forward scattering interactions ${\bf M}_{jk}$ in ${\cal H}^0_{\rm SLL}$.  When $\Delta^{\{m_p,n_p\}} > 2$ the operator $O^{\{m_p,n_p\}}$ is irrelevant and $v^{\{m_p,n_p\}}$ does not destabilize the gapless sliding Luttinger liquid fixed point.  When $\Delta^{\{m_p,n_p\}} < 2$, $O^{\{m_p,n_p\}}$ is relevant and $v^{\{m_p,n_p\}}$ grows at low energy, destabilizing the sliding Luttinger liquid.
%The scaling dimension $\Delta^{\{m_p,n_p\}}$ depends on the forward scattering interactions %${\bf M}_{jk}$.
In principle ${\bf M}_{jk}$ can be parameterized given an underlying model of the electron-electron interactions.  However, ${\bf M}_{jk}$ may be renormalized by irrelevant and/or momentum non conserving operators, so it may not resemble the bare interactions.  Here we follow the approach of Ref. \onlinecite{kml} and assume that ${\bf M}_{jk}$ have values such that a particular operator (or set of operators) $O^{\{m_p,n_p\}}$ is relevant.  Our object is to characterize the resulting non trivial strong coupling phases.  There are special values of ${\bf M}_{jk}$ that lead to particularly simple boson Hamiltonians that can be solved exactly.  These solvable points provide a powerful way to characterize the resulting strong coupling phases.

As shown in Ref. \onlinecite{kml} a number of non trivial 2D phases can be analyzed using this approach, including Abelian fractional quantum Hall states, superconductors and crystals of electrons, quasiparticles or vortices.  In particular, Abelian quantum Hall states are described by a single relevant operator $\{m_p,n_p\}$ satisfying
$\sum_p n_p \ne 0$.  From \eqref{nu} and \eqref{momentum conservation} this corresponds to a filling factor
\begin{equation}
\nu = 2 {{\sum_p p n_p}\over {\sum_p m_p}}.
\label{numn}
\end{equation}
In Section II.C we will review this construction for the Laughlin states and the Abelian hierarchy states.  But first, we will show that the coupled wire construction can also be straightforwardly applied to systems of bosons.

\subsection{Coupled Wire Construction for Bosons}

We now consider coupled wires of one dimensional bosons.  The low energy excitations of a single
wire can be described by ``bosonizing the bosons", to express them in terms of a slowly varying phase $\varphi(x)$ and a conjugate density variable $\theta(x)$ satisfying \eqref{thetaphicomm}.  The density fluctuations have important contributions near wavevectors $q_n= 2\pi n \bar\rho$ that are multiples of the average 1D density $\bar\rho$.
\begin{equation}
\rho(x) = \bar\rho + \sum_n \rho_n(x)
\end{equation}
As with the fermions, the long wavelength density fluctuation is
$\rho_0(x) = \partial_x\theta(x)/\pi$.
The density wave at $q\sim 2\pi \bar\rho n$ is,
\begin{equation}
\rho_n(x) \propto e^{in ( 2k_F x + 2\theta(x) )}.
\end{equation}
Here and in the following we will denote the 1D density $\bar\rho$ in terms of ``$2k_F$"$\equiv 2\pi \bar\rho$.  This allows us to proceed analogously with the fermions and use formulas \eqref{nu} and \eqref{momentum conservation} for the filling factor.

The Hamiltonian for bosonic wires coupled only by long wavelength interactions
has exactly the same form as ${\cal H}^0_{\rm SLL}$.  The only difference is that the non interacting Pauli
compressibility term ${\cal H}^0_{\rm non \ int.}$ is absent.  Tunneling a boson between wire $j$ and $j+p$ in the presence of a magnetic field is described by the operator
\begin{equation}
\Phi_{j+p}^\dagger(x) \Phi_j(x) e^{i b p x},
\end{equation}
where $\Phi^\dagger_j(x) \propto \exp i\varphi_j(x)$ is the boson creation operator.  Due to interactions this process can involve scattering from the $2k_F n$ density fluctuations of the bosons.
The most general coupling term thus has the form
\begin{equation}
{\cal O}^{\{m_p,n_p\}}_j =  c R^{\{m_p,n_p\}}\exp i\left( \sum_p n_p \varphi_{j+p}+m_p\theta_{j+p}\right),
\end{equation}
where $R$ is given in \eqref{R(x)}.
This is almost identical to the inter channel scattering terms for fermions.  The only differences are the absence of Klein factors and the constraints on the allowed values of $\{m_p,n_p\}$.
Charge and momentum conservation still requires (\ref{charge conservation}, \ref{momentum conservation}), but unlike for fermions, where $m_p$ and $n_p$ obey \eqref{fermionmn}, the corresponding constraint for bosons is
\begin{equation}
m_p = 0 \ {\rm mod}\ 2.
\label{bosonmn}
\end{equation}
The analysis of bosonic states then follows in exactly the same manner as fermionic states, as described in
Eqs. \eqref{h0sll+v}-\eqref{numn}.

\subsection{Laughlin States $\nu=1/m$}

Here we will examine the coupled wire construction for the Laughlin states in some detail.  We include the details here because the Laughlin states provide the simplest non trivial application of the coupled wire construction.  We begin by introducing the relevant interaction term, and then characterize the bulk quasiparticles and edge states.

\subsubsection{Tunneling Hamiltonian}

The Laughlin sequence of quantum Hall states at filling $\nu = 1/m$ are characterized by the correlated tunneling operators involving two neighboring wires.  The relevant operator is associated
with a link $\ell \equiv j+1/2$ between wires $j$ and $j+1$,
\begin{equation}
{\cal O}_{\ell} = \exp i \left[\varphi_{j} - \varphi_{j+1} + m(\theta_j + \theta_{j+1})\right].
\label{OLaughlin}
\end{equation}

Using \eqref{numn} it can readily be seen that this term is allowed for magnetic fields $b$ corresponding to $\nu = 1/m$.  Moreover, from \eqref{fermionmn} and \eqref{bosonmn}, it is clear that $m$ odd (even) corresponds to a fermionic (bosonic) state, as is expected for the Laughlin state.  In Eq. \ref{OLaughlin}  we have suppressed the Klein factors, which are necessary for fermions.  They are treated carefully in Appendix A.  This term is represented schematically in Fig. 2a.  The notation for this diagram is slightly different from the one used in Ref \onlinecite{kml}.  The vertical arrows describe the tunneling of charge between the wires (represented by $\phi_{j+1}-\phi_j$), while the circular arrows describe backscattering within a wire (represented by $\theta_{j,j+1}$).  Note that the number of $\theta$'s is constrained by \eqref{fermionmn} or \eqref{bosonmn}.

\begin{figure}
\includegraphics[width=3in]{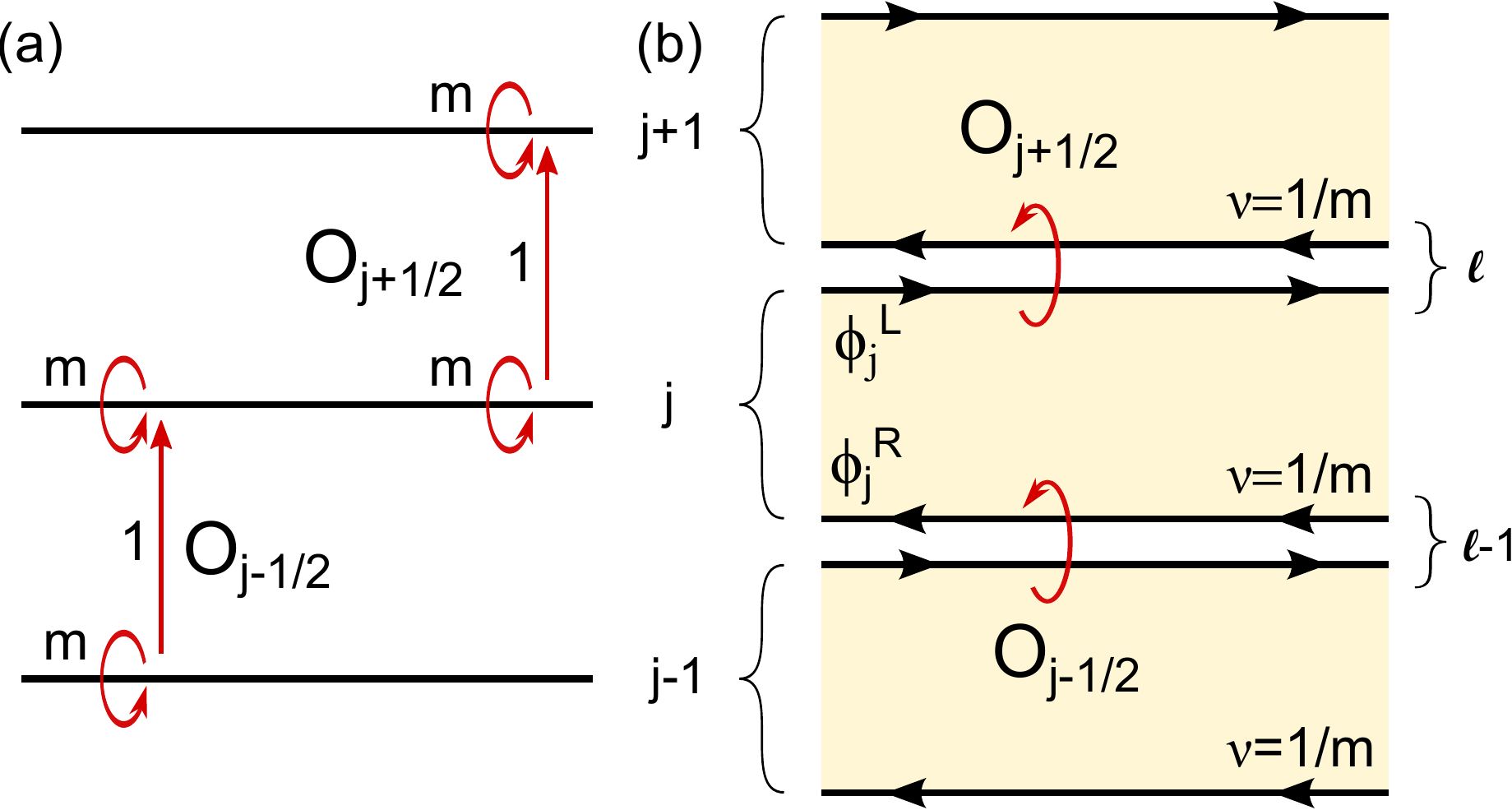}
\caption{(a) Schematic representation of the correlated tunneling processes in \eqref{OLaughlin} that lead to the Laughlin state.  Vertical arrows represent $\phi$ and circular arrows represent $\theta$.  After the change of variables \eqref{phirl}, the model describes strips of $\nu = 1/m$ quantum Hall fluid coupled by tunneling electrons between their edge states. }
\label{Fig 2}
\end{figure}

The tunneling operators defined above have the special property that they all commute with eachother.  In particular,
$[{\cal O}_{j+1/2},{\cal O}_{j-1/2}]=0$.  This means that the components from wire $j$ in those two operators
commute with one another.  This invites us to introduce right and left moving chiral fields on wire $j$ that distinguish the two contributions
to ${\cal O}_{j\pm 1/2}$.  We thus write
\begin{eqnarray}
\tilde\phi_j^R &=& \varphi_j + m \theta_j \nonumber\\
\tilde\phi_j^L &=& \varphi_j - m \theta_j.
\label{phirl}
\end{eqnarray}
The decoupling can be explicitly seen from the commutation algebra,
\begin{equation}
\left[\partial_x\tilde\phi_j^p(x),\tilde\phi_{j'}^{p'}(x')\right] = 2\pi i m p \delta_{pp'}\delta_{jj'}\delta_{xx'}.
\label{phirl tilde commutator}
\end{equation}
%\begin{eqnarray}
%\left[\partial_x\phi_j^R(x),\phi_{j'}^R(x')\right] &=& 2\pi i m \delta_{jj'} \delta_{xx'}\\
%\left[\partial_x\phi_j^L(x),\phi_{j'}^L(x')\right] &=& -2\pi i m\delta_{jj'} \delta_{xx'}\\
%\left[\partial_x\phi_j^R(x),\phi_{j'}^L(x')\right] &=& 0.

%\end{eqnarray}
The interaction term is now ${\cal O}_{j+1/2} = \exp i(\tilde\phi_{j+1}^R - \tilde\phi_j^L)$.
The charge density is
$\rho_j = (\partial_x\tilde\phi_j^R - \partial_x\tilde\phi_j^L)/(2\pi m)$.

It is also convenient to introduce new density and phase variables defined on the links $\ell=j+1/2$ between wires,
\begin{eqnarray}
\tilde\theta_{\ell} &=& (\tilde\phi_{j}^R - \tilde\phi_{j+1}^L)/2 \nonumber\\
\tilde\varphi_{\ell} &=& (\tilde\phi_{j}^R + \tilde\phi_{j+1}^L)/2.
\label{thetaphi ell}
\end{eqnarray}
These satisfy
$
[\partial_x\tilde\theta_{\ell}(x),\tilde\theta_{\ell'}(x')] =
[\partial_x\tilde\varphi_{\ell}(x),\tilde\varphi_{\ell'}(x')] = 0$ and
\begin{equation}
[\partial_x\tilde\theta_{\ell}(x),\tilde\varphi_{\ell'}(x')] = i\pi m\delta_{\ell\ell'}\delta_{xx'}.
\label{thetaphitildecomm}
\end{equation}
The charge density associated with the link $\ell$ can be written
\begin{equation}
\tilde \rho_\ell = \partial_x\tilde\theta_\ell/(m\pi).
\label{rho}
\end{equation}

In terms of the new variables, the Hamiltonian becomes,
\begin{equation}
{\cal H} = \tilde{\cal H}^0_{\rm SLL}[\tilde\theta,\tilde\varphi] + \sum_\ell \int dx v \cos 2\tilde\theta_{\ell},
\label{h0+vcos}
\end{equation}
where without loss of generality we have assumed $v$ is real.  As shown in Appendix A, there is no Klein factor, provided the zero momentum component of $\tilde\theta_\ell$ is correctly defined.

Provided the forward scattering interactions defining
${\cal H}^0_{\rm SLL}$ are such that $v$ is relevant, the system will flow at low energy to a gapped
phase in which $\tilde\theta_\ell$ is localized in a well of the cosine potential.  As argued in Ref. \onlinecite{kml}, it is always possible
to find such interactions.  In particular, consider a simple interaction such that $\tilde{\cal H}^0_{\rm SLL}$ has the decoupled form
\begin{equation}
\tilde {\cal H}^0_{\rm SLL} = {v_0\over {2\pi}} \sum_\ell \int dx
 \left( {1\over g} (\partial_x\tilde\theta_\ell)^2 +   g (\partial_x\tilde\varphi_\ell)^2\right) .
\end{equation}
The scaling dimension of $\cos 2\tilde\theta_\ell$ is $\Delta = m g$. It follows that for $g < 2/m$, $v$ is relevant.   It should be emphasized that ${\cal H}^0_{\rm SLL}$ can be expressed in terms of the original fermion operators, which includes a specific four fermion forward scattering interaction.  For special values of $g$ this model can be solved exactly.  In the limit $g \rightarrow 0$, the variable $\tilde\theta_\ell$ becomes a stiff classical variable, so that the approximation of replacing $-\cos2\tilde\theta_\ell$ by $2\tilde\theta_\ell^2$ becomes exact.
For larger $g$, we rely on our understanding that $g$ is renormalized downward by $v$, so that
at $\tilde\theta$ stiffens at low energy.
For $g=1/m$ there is another exact solution because it is possible to define new variables such that the Hamiltonian has precisely the form of \eqref{h0 free fermion}.  The problem can then be refermionized and expressed in terms of non interacting fermions which have a single particle energy gap.  We will not dwell on these exact solutions any further in this paper.  We will be content with our understanding that any $g<2/m$ leads to a gapped state.

The gapped phase is the Laughlin state\cite{laughlin}.  This can be seen by examining the quasiparticle excitations and the edge states.

\subsubsection{Bulk Quasiparticles}

Quasiparticles occur when $\tilde\theta_\ell(x)$ has a kink where it jumps by $\pi$.  From \eqref{rho} it can be seen that such a kink is associated with  a charge $e/m$.  This makes the charge fractionalization in the fractional quantum Hall effect appear similar to the fractionalization that occurs in the one dimensional Su Schrieffer Heeger (SSH) model\cite{ssh}.  However, there is a fundamental difference.  The solitons in the SSH model occur at domain walls separating {\it physically distinct states}.  This prevents solitons from hopping between wires via a local operator.  In contrast, the states characterized by $\tilde\theta_\ell$ and $\tilde\theta_\ell+\pi$ are physically equivalent.  They are related by a gauge transformation in which, say, $\varphi_j \rightarrow \varphi_j+2\pi$, which takes $\tilde\theta_{j\pm 1/2}$ to $\tilde\theta_{j\pm 1/2} \mp \pi$.  This allows quasiparticles to hop via a local operator without the nonlocal string.
Though SSH solitons and Laughlin quasiparticles are distinct, they become equivalent on a cylinder with finite circumference.  The Tao Thouless ``thin torus" limit\cite{taothouless} can be described the extreme case in which the ``cylinder" consists of a single wire with electron tunneling ``around" the cylinder.  In this case, our theory maps precisely to an $m$ state version of the SSH model.

The local operator that hops quasiparticles between links $j+1/2$ and $j-1/2$ is simply the backscattering of a {\it bare} electron on wire $j$, $\psi_j^{L\dagger}\psi_j^R$ (or equivalently for bosons the $2k_F$ density operator).  Using the transformations \eqref{phirl} and \eqref{thetaphi ell} it is straightforward to show that
\begin{eqnarray}
\chi_j(x') &\equiv& \psi_j^{L\dagger}\psi_j^R \nonumber \\
&=& e^{2i\theta_j(x')} = e^{i(\tilde\phi_j^R(x') - \tilde\phi_j^L(x'))/m}
\label{chij}\\
&=& e^{i(\tilde\varphi_{j+1/2}-\tilde\varphi_{j-1/2} +\tilde\theta_{j+1/2} + \tilde\theta_{j-1/2})/m}.\nonumber
\end{eqnarray}
From \eqref{thetaphitildecomm} it can be seen that this operator takes $\partial_x\tilde\theta_{j\pm 1/2}$ to $\partial_x\tilde\theta_{j\pm 1/2} \mp \pi \delta(x-x')$, transferring a quasiparticle from $j+1/2$ to $j-1/2$.   The operator that transfers a quasiparticle along wire from $x$ to $x'$ along wire link $\ell$ is
\begin{equation}
\rho_\ell(x,x') = e^{i(\tilde\varphi_\ell(x)-\tilde\varphi_\ell(x'))/m} = e^{i\int_{x'}^x dx \partial_x\tilde\varphi_\ell/m},
\label{rholxx'}
\end{equation}
which can also be expressed in terms of the bare electron densities and currents.

A quasiparticle operator may be defined as
\begin{equation}
\Psi^{R/L}_{QP,j+1/2}(x) = e^{i \tilde\phi_{j/j+1}^{R/L}/m} = e^{i (\tilde\varphi_{j+1/2} +/- \tilde\theta_{j+1/2})/m}
\end{equation}
In the bulk, since $\tilde\theta$ is gapped, we have
$
\Psi^{R}_{QP,\ell} = \Psi^{L}_{QP,\ell} e^{2 i \langle \tilde\theta_\ell \rangle}
$.
Of course, since $\Psi^{R/L}_{QP,\ell}$ can not be locally built out of bare electron operators, it is not by itself a physical operator.  However, the operator that transfers a quasiparticle from one location to another can be built from a string of local operators like \eqref{chij} and \eqref{rholxx'}.  This allows the fractional statistics of the quasiparticles to be seen quite simply.

To move a quasiparticle from $x_1$ to $x_2$ on link $\ell_1$ and then to $x_2$ on $\ell_2$, use the operator
\begin{equation}
\rho_{\ell_1}(x_1,x_2)\prod_{j=\ell_1+1/2}^{\ell_2-1/2} \chi_j(x_2) .
\label{qpmove}
\end{equation}
Since $\tilde\theta$ is gapped, this can be written
\begin{equation}
\Psi^{R\dagger}_{QP,\ell_2}(x_2)\Psi^{L}_{QP,\ell_1}(x_1) \prod_{\ell=\ell_1+1}^{\ell_2-1} e^{2 i \langle\tilde\theta_{\ell}(x_2)\rangle/m}
\end{equation}
The string of $\langle \tilde\theta\rangle$ is responsible for the fractional statistics.

Consider moving a quasiparticle through a closed loop.  The operator that takes a quasiparticle around the rectangle formed by $x_1$, $x_2$, $\ell_1$ and $\ell_2$ can be constructed by doing \eqref{qpmove} twice, which eliminates the quasiparticle operators.  This then gives a phase
\begin{equation}
%\Pi_\square(x_1,x_2,\ell_1,\ell_2) &=&
\prod_{\ell=\ell_1+1}^{\ell_2-1} e^{2 i(\langle\tilde\theta_{\ell}(x_2)\rangle - \langle\tilde\theta_{\ell}(x_1)\rangle)/m}
= e^{2\pi i N_{QP}/m},
\end{equation}
where $N_{QP}$ is the number of quasiparticles enclosed by the rectangle.  Here we have used the fact that  $(\langle\tilde\theta_\ell(x_1)\rangle - \langle\tilde\theta_\ell(x_2)\rangle)/2\pi$ simply counts the number of quasiparticles on link $\ell$ between $x_1$ and $x_2$.

\subsubsection{Edge States}

For a finite array of wires with open boundary conditions, the edge states are apparent, since there are extra chiral modes left over on the first and last wire.  From \eqref{phirl tilde commutator}, it can be seen that these modes have precisely the chiral Luttinger liquid structure of $\nu = 1/m$ edge states\cite{wenll}.
\begin{equation}
{\cal H}_{\rm edge} = {m v_0\over{4 \pi}}(\partial_x\tilde\phi_1^L)^2
\end{equation}
with $[\partial_x\phi_1^R(x),\phi_1^R(x')] = 2\pi i \delta_{xx'}$.
The electron operator on the $j=1$ edge is
\begin{equation}
\Psi^e_1 = e^{i\tilde\phi_1^L}.
\end{equation}
It is straightforward to show that this operator has the expected dimension $\Delta = m/2$, characteristic of the chiral Luttinger liquid.

One can view the change of variables \eqref{phirl} as a transformation between a sliding Luttinger liquid built out of bare electrons and a sliding Luttinger liquid built out of $\nu = 1/m$ edge states.  The correlated tunneling term for the bare electrons becomes the electron tunneling operator coupling the edge states.  The array of wires then becomes an array of strips of $\nu=1/m$ quantum Hall fluid coupled by electron tunneling, as shown in Fig. \ref{Fig 2}.  When the electron tunneling is relevant the strips merge to form a single bulk $\nu=1/m$ fluid, leaving behind chiral modes at the edge.

The quasiparticle operator at the $j=1$ edge is
\begin{equation}
\Psi^{L}_{QP,1} = e^{i\phi_1^L/m}.
\end{equation}
As discussed above, since $\Psi_{QP}$ can not be made out of bare electron operators, it is not by itself a physical operator.  However, quasiparticle tunneling from the top to the bottom edge can be built from a string of backscattering operators \eqref{chij}.  When the gapped bulk degrees of freedom are integrated out, this string of operators becomes
\begin{equation}
\prod_{j=1}^N \chi_j \sim \Psi^{L\dagger}_{QP,1} \Psi^{R}_{QP,N}.
\end{equation}

%\subsection{Quasiparticle Statistics and Ground State Degeneracy}

\subsection{Hierarchy States}

In this section we show how the coupled wire construction describes hierarchical Abelian fractional quantum Hall states\cite{haldanehierarchy,halperin}.  We restrict ourselves to second level states, which are characterized by a $2\times 2$ $K$ matrix\cite{wenzee}.  Generalization to higher levels is straightforward.

\begin{figure}
\includegraphics[width=3in]{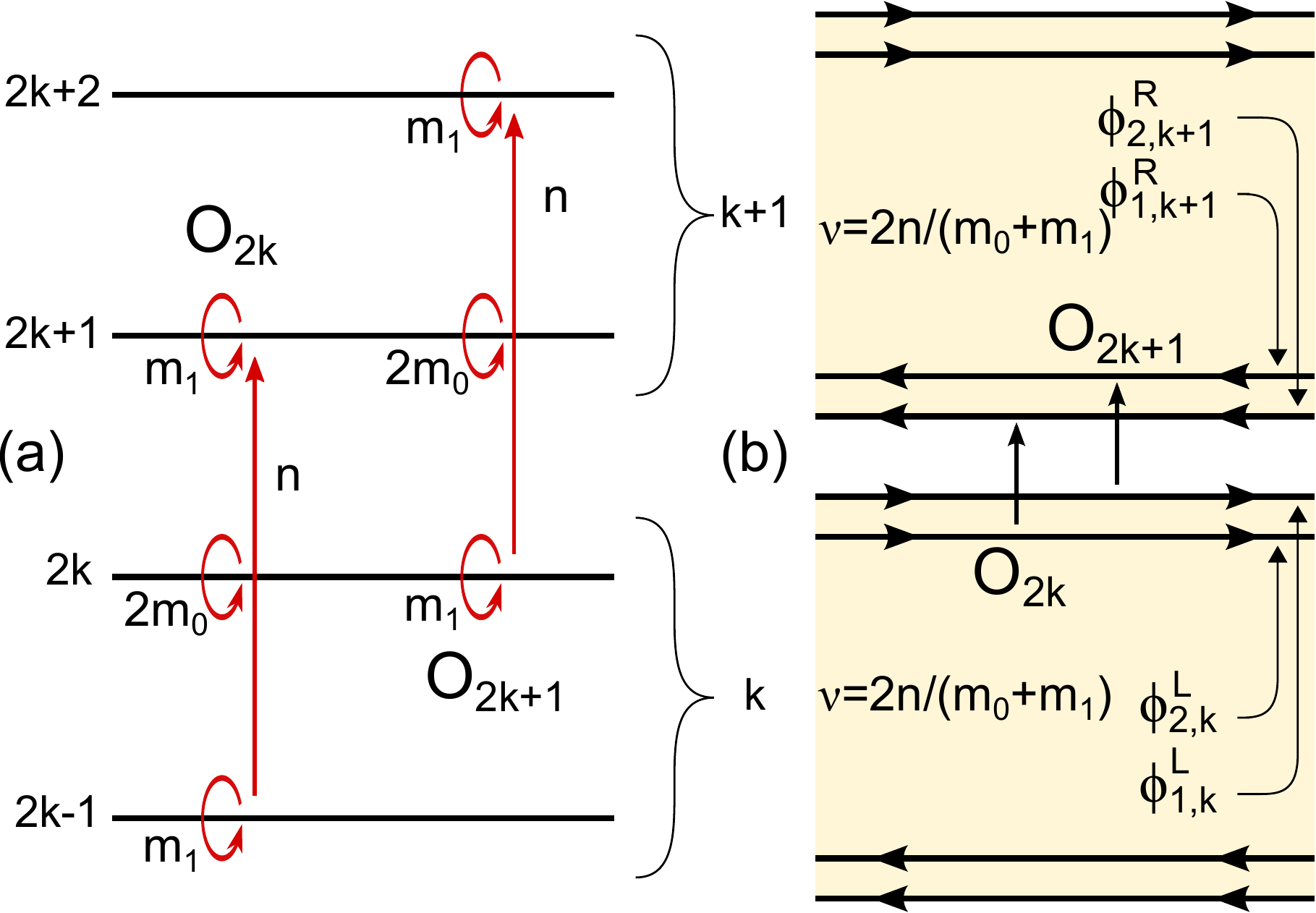}
\caption{(a) Schematic of tunneling processes in \eqref{Ohierarchy} that lead to 2nd level Abelian hierarchy states.  (b) After the transformation \eqref{phirl hierarchy} the model describes coupled
strips of $\nu=2n/(m_0+m_1)$ quantum Hall state coupled by tunneling electrons between the two channels of edge states.}
\label{Fig 3}
\end{figure}

2nd level hierarchy states arise from an interaction term that involve three coupled wires.  A generic term is shown in Fig. \ref{Fig 3}, and can be described by the operator
\begin{equation}
{\cal O}_j =  \exp i\left[n(\varphi_{j-1} - \varphi_{j+1})+ 2 m_0\theta_j + m_1(\theta_{j+1}+\theta_{j-1}) \right].
\label{Ohierarchy}
\end{equation}
Here $n$ and $m_0$ are any integers, while $m_1$ ($m_1+n$) is an even integer for bosons (fermions).  Again, we defer discussion of the Klein factors to Appendix A.  From \ref{numn}, this interaction conserves momentum at a filling factor
\begin{equation}
\nu = {2n\over{m_0+m_1}}
\end{equation}
This set of states corresponds to the standard Haldane-Halperin hierarchy states at filling $(p_0 + 1/p_1)^{-1}$ ($p_0$ is even (odd) for bosons (fermions) and $p_1$ is even) for the choice,
$n = p_1/2$, $m_0 = p_0 p_1/2$ and $m_1 = p_0 p_1/2 +1$.

To analyze this state, we group the wires into pairs $j = 2k$ and $j=2k+1$.  Pair $k$ is connected to pair $k+1$ by two tunneling terms, ${\cal O}_{2k}$ and ${\cal O}_{2k+1}$.  As in \eqref{phirl} we define new variables that decouple right and left moving modes on the pairs of wires.
\begin{eqnarray}
\tilde\phi_{k,1}^R  &=& n \varphi_{2k-1} + m_1 \theta_{2k-1} + 2 m_0 \theta_{2k} \nonumber\\
\tilde\phi_{k,1}^L  &=& n \varphi_{2k-1} - m_1 \theta_{2k-1} \nonumber \\
\tilde\phi_{k,2}^R  &=& n \varphi_{2k} + m_1 \theta_{2k}
\label{phirl hierarchy}\\
\tilde\phi_{k,2}^L  &=& n \varphi_{2k} - m_1 \theta_{2k} - 2 m_0 \theta_{2k-1} \nonumber.
\end{eqnarray}
The new fields obey the commutation algebra
\begin{equation}
\left[\partial_x\tilde\phi_{k,a}^p(x),\tilde\phi_{k',b}^{p'}(x') \right] =  2\pi i  p \delta_{pp'}\delta_{kk'}K_{ab} \delta_{xx'}
\label{phirl hierarchy commutator}
\end{equation}
where the $K$ matrix is
\begin{equation}
K_{ab} = n \left(\begin{array}{cc}
    m_1 & m_0 \\ m_0 & m_1
\end{array} \right).
\label{k matrix}
\end{equation}
The charge density is
\begin{equation}
\rho_{k} = \sum_{a} t_a \partial_x (\phi^R_{k,a}-\phi^L_{k,a}) /2\pi
\end{equation}
with
\begin{equation}
t_a = {1\over{m_0+m_1}}
\left( \begin{array}{c} 1\\ 1  \end{array}\right).
\end{equation}

We next define variables on links $\ell = k+1/2$,
\begin{eqnarray}
\tilde\theta_{a,\ell=k+1/2} &=& (\phi_{k,a}^R - \phi_{k+1,a}^L )/2\\
\tilde\varphi_{a,\ell=k+1/2} &=& (\phi_{k,a}^R + \phi_{k+1,a}^L )/2.
\end{eqnarray}
These satisfy, $[\partial_x\tilde\theta_{\ell,a},\tilde\theta_{\ell',b}]=[\partial_x\tilde\varphi_{\ell,a},\tilde\varphi_{\ell',b}]=0$ and
\begin{equation}
[\partial_x\tilde\theta_{\ell,a},\tilde\varphi_{\ell',b}] = i\pi K_{ab}\delta_{\ell\ell'}\delta_{xx'}.
\end{equation}
In terms of these new variables, the Hamiltonian in the presence of the correlated $n$-electron tunneling operators becomes
\begin{equation}
{\cal H} = {\cal H}^0_{\rm SLL}[\tilde\theta_{\ell,a},\tilde\varphi_{\ell,a}] + \sum_\ell \int dx v\left( \cos 2\tilde\theta_{\ell,1} + \cos 2\tilde\theta_{\ell,2} \right).
\label{h0sll+vcos h}
\end{equation}
If $v$ flows to strong coupling, we have a gapped bulk, describing a $\nu= 2n/(m_0+m_1)$
quantum Hall fluid characterized by the $K$ matrix \eqref{k matrix}.   As in Section II.C.3, this can be interpreted as quantum Hall strips with edge states coupled by the charge $n e$ tunneling operators
\begin{equation}
\Psi^{ne,R/L}_{k,a} = e^{i\phi_{k,a}^{R/L}}.
\end{equation}
Quasiparticles, given by $\pi$ kinks in $\tilde\theta_{\ell,1}$ or $\tilde\theta_{\ell,2}$ are created by
\begin{equation}
\Psi^{R/L}_{QP,a,k+1/2} = e^{i \sum_b K^{-1}_{ab} \phi_{k/k+1,b}^{R/L}}.
\end{equation}
They have charge $e/(m_0+m_1)$.
The bare electron backscattering operator corresponds to quasiparticle tunneling,
\begin{equation}
\chi_{k,a} \equiv e^{2i\theta_{2k-2+a}} = \Psi^{L\dagger}_{QP,a,k-1/2} \Psi^{R}_{QP,a,k+1/2}.
\end{equation}

At the edge of a semi infinite system, where will be two chiral modes left over described by $\tilde\phi^R_{1,a}$.  From \eqref{phirl hierarchy commutator} it can be seen that these give precisely the chiral Luttinger liquid edge states characterized by the K matrix \eqref{k matrix}.

\section{Moore Read State}

We now generalize the coupled wire construction to describe the Moore Read state\cite{mr}.  Our approach was motivated by the observation by Fradkin, Nayak and Shoutens\cite{fradkin} that the Moore Read state for bosons at filling $\nu=1$ has a simple interpretation in terms of two coupled copies of bosons at $\nu=1/2$.  Each copy is described by a $SU(2)_1$ Chern Simons theory, and the coupling between them introduces the symmetry breaking $SU(2)_1\times SU(2)_1 \rightarrow SU(2)_2$\cite{fradkin2}.

We therefore first consider the problem of coupled wires of bosons at filling $\nu=1$, where the bosons on each wire have two flavors, each at $\nu=1/2$.  The allowed boson tunneling and backscattering terms in our construction have a simple representation in the low energy bosonized theory.  Moreover, by fermionizing the bosons, the Majorana fermions associated with the Moore Read state\cite{readgreen} emerge naturally.

There is a special set of values for the interactions in which the problem is particularly simple.  In this case, the Hilbert space associated with the two right (left) moving chiral modes on each wire decouples into two sectors.  One of the sectors is coupled to the corresponding sector of the left (right) moving modes on the {\it same} wire, while the other sector is coupled to the corresponding sector of the left (right) moving modes on the {\it neighboring} wire.  Both of these couplings introduce gaps, but the two sectors are gapped in ``opposite directions".  This gives a kind of hybrid between the insulating phase, in which all chiral modes are paired on the same wire, and the quantum Hall states, in which all the chiral modes are coupled on neighboring wires.  What gets left behind on the edge is a fraction of the original chiral modes.

This fractionalization of the original chiral modes described mathematically in terms of the coset construction in conformal field theory\cite{coset,cft}.  The original pair of chiral modes are described by a  $SU(2)_1 \times SU(2)_1$ theory with central charge $c=2$. These modes decompose into three sectors:
$SU(2)_1 \times SU(2)_1/SU(2)_2$, $SU(2)_2/U(1)$ and $U(1)$ with $c= 1/2$, $1/2$ and $1$, respectively.
The independent sectors are then gapped ``in different directions".   We will describe this construction in Section III.B. This, in effect, gives a concrete and somewhat more explicit implementation of the Fradkin, Nayak, Shoutens construction.

After establishing the Moore Read state for $\nu=1$ bosons, we will go on to generalize our construction to account for fermions, and the $q$-pfaffian state at filling $\nu = 1/(q+1)$, where $q$ is even (odd) for bosons (fermions).

\subsection{Bosons at $\nu=1$}

\begin{figure}
\includegraphics[width=3in]{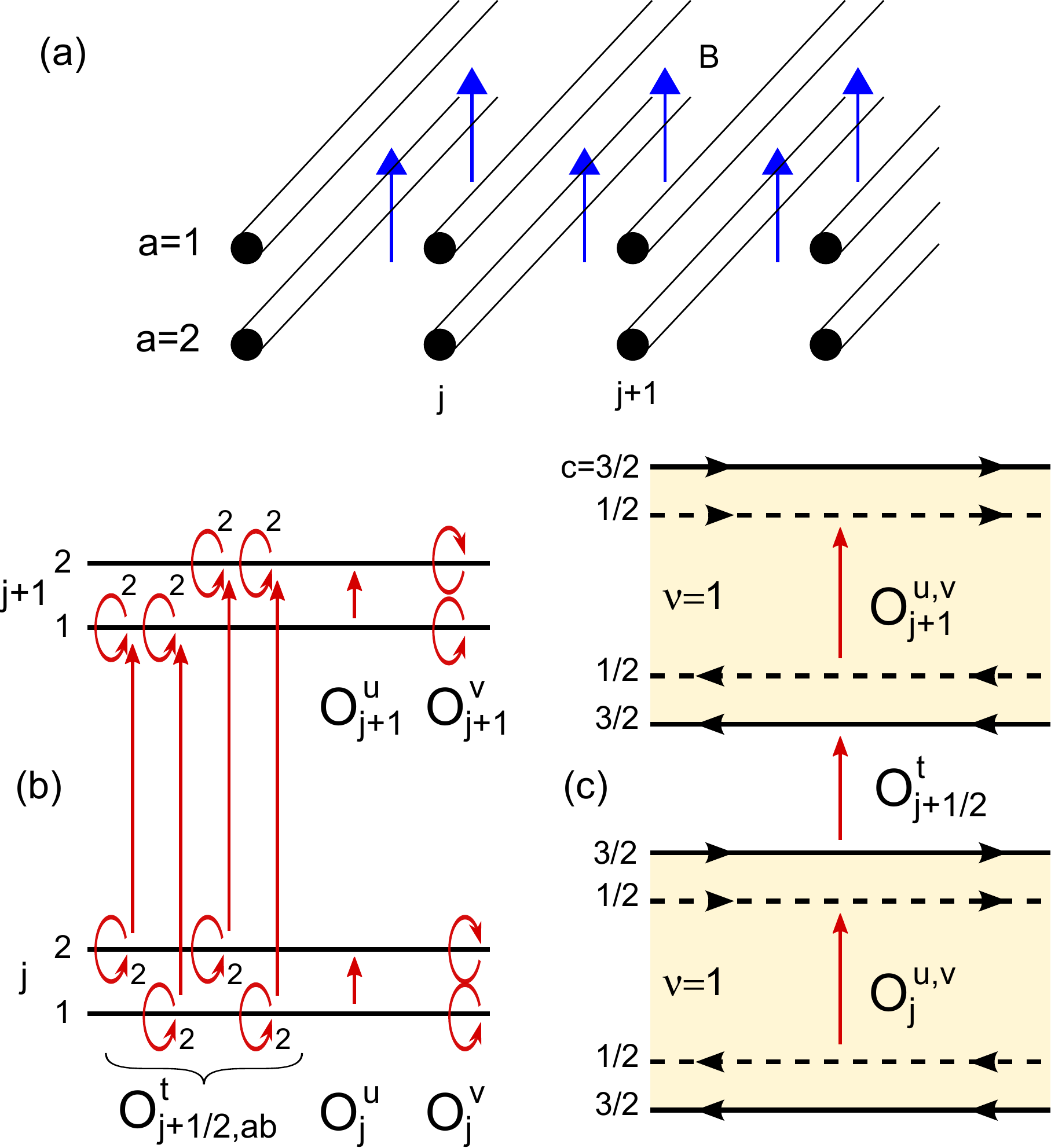}
\caption{(a) A two component coupled wire model, viewed as a bilayer.  (b) Schematic of tunneling processes in (\ref{Ot},\ref{Ou},\ref{Ov}) that
lead to the Moore Read state for bosons at filling $\nu=1$. (b) After the
transformations (\ref{chiral mr},\ref{chiral mr rhosigma},\ref{fermionize}) the model describes coupled strips of $\nu=1$ Moore Read state coupled by tunneling
electrons between the edge states, which are characterized by a $c=1$ chiral charge mode and a $c=1/2$ chiral Majorana fermion mode.}
\label{Fig 4}
\end{figure}

We begin with a Hamiltonian ${\cal H} = {\cal H}^0_{\rm SLL} +V$ describing coupled wires of two component bosons, which can be viewed as a double layer system, as in Fig. \ref{Fig 4}a.  Each component has a density corresponding to filling $\nu=1/2$. ${\cal H}^0$ has the same form as \eqref{h0sll}, except now each wire has two bosons, $\theta_{j,a}$ and $\varphi_{j,a}$, for $a=1,2$, which satisfy
\begin{equation}
[\partial_x\theta_{j,a}(x),\varphi_{j',a'}(x')] = i\pi\delta_{jj'}\delta_{aa'}\delta_{xx'}.
\label{thetaphi mr commutator}
\end{equation}
The interaction terms $V$ consist of boson tunneling and backscattering operators that are
consistent with momentum conservation.  We consider three such terms, depicted in Fig. \ref{Fig 4}.
\begin{equation}
V = \sum_j \int dx\left( \sum_{ab=1}^2 t_{ab} {\cal O}^t_{j,ab}
+ u {\cal O}^u_j + v {\cal O}^v_j\right) + h.c.
\end{equation}
The first term involves coupling between channel a on one wire and channel b on the neighboring wire.
\begin{equation}
{\cal O}^t_{j;ab} = e^{i\left (\varphi_{j,a} - \varphi_{j+1,b} + 2(\theta_{j,a}+\theta_{j+1,b})\right)}.
\label{Ot}
\end{equation}
This term is similar to \eqref{OLaughlin}.  The coefficient 2 of the $\theta$ terms is fixed by the filling factor.  In addition, there are allowed terms that couple the two channels on a single wire.  These include a Josephson like coupling between the two channels,
\begin{equation}
{\cal O}^u_{j} = e^{i\left( \varphi_{j,1} - \varphi_{j,2} \right)},
\label{Ou}
\end{equation}
as well as an interaction that locks the ``$2k_F$" densities of the two channels,
\begin{equation}
{\cal O}^v_{j} = e^{i\left( 2\theta_{j,1} - 2\theta_{j,2} \right)}.
\label{Ov}
\end{equation}
These three terms (as well as combinations of them) are the only allowed interaction terms at $\nu=1/2$ that include up to first neighbor coupling.

It is now useful to introduce right and left moving chiral fields,
\begin{eqnarray}
\tilde\phi_{j,a}^R &=& \varphi_{j,a} + 2\theta_{j,a} \nonumber\\
\tilde\phi_{j,b}^L &=& \varphi_{j,a} - 2\theta_{j,a},
\label{chiral mr}
\end{eqnarray}
as well as ``charge" and ``spin" fields,
\begin{eqnarray}
\tilde\phi_{j,\rho}^p &=& (\tilde\phi_{j,1}^p + \tilde\phi_{j,2}^p)/2,\nonumber \\
\tilde\phi_{j,\sigma}^p &=& (\tilde\phi_{j,1}^p - \tilde\phi_{j,2}^p)/2.
\label{chiral mr rhosigma}
\end{eqnarray}
The latter fields satisfy
\begin{equation}
\left[\partial_x\tilde\phi_{j,\mu}^p(x),\tilde\phi_{j',\mu'}^{p'}(x')\right] = 2\pi i p \delta_{pp'}\delta_{\mu\mu'} \delta_{jj'} \delta_{xx'}
\label{phirl mr commutator}
\end{equation}
for $\mu = \rho, \sigma$.

For simplicity, we will first focus on the case in which $t_{ab} = t$, independent of $a$ and $b$, and $t, u$ and $v$ are real.  We
will comment on the more general case later.  In this case, in terms of the new variables, we have
\begin{eqnarray}
\sum_{ab} i t_{ab} {\cal O}^t_{j,ab} + h.c. = \hspace{1.5in} \nonumber\\
8t \cos(\tilde\phi^R_{j,\rho}-\tilde\phi^L_{j+1,\rho})) \cos \tilde\phi^R_{j,\sigma} \cos \tilde\phi^L_{j+1,\sigma}.
\label{Ot+hc}
\end{eqnarray}
and
\begin{eqnarray}
u {\cal O}^u_j + h.c. = 2u \cos(\tilde\phi^R_{j,\sigma} + \tilde\phi^L_{j,\sigma})\hspace{.8in} \nonumber\\%= 2 u \cos 2\varphi_{j,\sigma}\\
= 2i u (\sin\tilde\phi^R_{j,\sigma} \sin\tilde\phi^L_{j,\sigma} - \cos\tilde\phi^R_{j,\sigma} \cos\tilde\phi^L_{j,\sigma} )
\label{Ou+hc}\\
v {\cal O}^v_j + h.c. = 2v \cos(\tilde\phi^R_{j,\sigma} - \tilde\phi^L_{j,\sigma})\hspace{.8in} \nonumber\\ %= 2 v \cos 2\theta_{j,\sigma}.
= 2i v (\sin\tilde\phi^R_{j,\sigma} \sin\tilde\phi^L_{j,\sigma} + \cos\tilde\phi^R_{j,\sigma} \cos\tilde\phi^L_{j,\sigma}  )
\label{Ov+hc}
\end{eqnarray}
Note that in passing between the first and second lines of \eqref{Ot+hc}-\eqref{Ov+hc} getting the factors of $i$ right requires care in splitting the exponential.  This is explained in Appendix A.3, where the zero momentum components of $\tilde\phi^{R/L}_{j,\sigma}$ are properly taken into account.

Since the interaction term $V$ is a sum of non commuting terms, analysis of this state is more complicated than it was for the Abelian quantum Hall states.  However, a tremendous simplification occurs when the forward scattering interactions in ${\cal H}^0_{\rm SLL}$ are such that $\tilde\phi^{R/L}_{j,\rho}$ and $\tilde\phi^{R/L}_{j,\sigma}$ are decoupled, and the Hamiltonian for $\tilde\phi^{R/L}_{j,\sigma}$ has the non interacting form, ${\cal H}^0_{\rm non int.}$ in
\eqref{h0 free fermion}.  In this case, the operator $\exp i\tilde\phi_{j,\sigma}$ has precisely the form of a bosonized Dirac fermion.  This allows us to fermionize, by writing
\begin{equation}
\psi^{R/L}_{j,\sigma} = \xi^{R/L}_{j,\sigma} + i \eta^{R/L}_{j,\sigma} = {\kappa_{j,\sigma}\over{\sqrt{2\pi x_c}}} e^{i\tilde\phi^{R/L}_{j,\sigma}}
\label{fermionize}
\end{equation}
where $\psi_\sigma$ is a Dirac fermion operator, and $\xi_{j,\sigma}$, $\eta_{j,\sigma}$ are Majorana fermion operators.   For the charge sector, we define
\begin{eqnarray}
\tilde\theta_{j+1/2,\rho} &=& (\tilde\phi^R_{j,\rho} - \tilde\phi^L_{j+1,\rho})/2 \nonumber\\
\tilde\varphi_{j+1/2,\sigma} &=& (\tilde\phi^R_{j,\rho} + \tilde\phi^L_{j+1,\rho})/2.
\end{eqnarray}
They satisfy $[\partial_x\tilde\theta_{\ell,\rho}(x),\tilde\theta_{\ell',\rho}(x')] = [\partial_x\tilde\varphi_{\ell,\rho}(x),\tilde\varphi_{\ell',\rho}(x')]=0$ and
\begin{equation}
[\partial_x\tilde\theta_{\ell,\rho}(x),\tilde\varphi_{\ell',\rho}(x')] = i\pi  \delta_{\ell\ell'}\delta_{xx'}.
\end{equation}

The Hamiltonian may now be written,
\begin{equation}
{\cal H} = {\cal H}_{\rm SLL}^0[\tilde\theta_{\ell,\rho},\tilde\varphi_{\ell,\rho}] + {\cal H}^0_{\rm M}[\eta_\sigma] + {\cal H}^0_{\rm M}[\xi_\sigma] + V
\end{equation}
where ${\cal H}_{\rm SLL}^0[\theta_{j,\rho},\phi_{j,\rho}]$ has the form \eqref{h0sll}.
For a special value of the interactions in the charge sector it is also possible to fermionize $\tilde\theta_{j,\rho}$ and $\tilde\varphi_{j,\rho}$, though that is not necessary for our purposes.
The free fermion Hamiltonian for the Majorana fermion $\eta^{R/L}_{j,\sigma}$ is
\begin{equation}
{\cal H}^0_{\rm M}[\eta_\sigma] =
\sum_j \int dx i\left( \eta^R_{j,\sigma} \partial_x \eta^R_{j,\sigma} - \eta^L_{j,\sigma} \partial_x \eta^L_{j,\sigma} \right),
\end{equation}
with a similar expression for ${\cal H}_0[\xi_{j,\sigma}]$.  The interaction term is
\begin{eqnarray}
V = \sum_j \int dx \left[ \tilde t \cos 2\tilde\theta_{j+1/2,\rho} i\xi_{j,\sigma}^R \xi_{j+1,\sigma}^L  + \right. \nonumber\\
\left.(\tilde u - \tilde v) i\xi_{j,\sigma}^R \xi_{j,\sigma}^L + (\tilde u+\tilde v) i\eta_{j,\sigma}^R \eta_{j,\sigma}^L \right].
\end{eqnarray}

\begin{figure}
\includegraphics[width=3.2in]{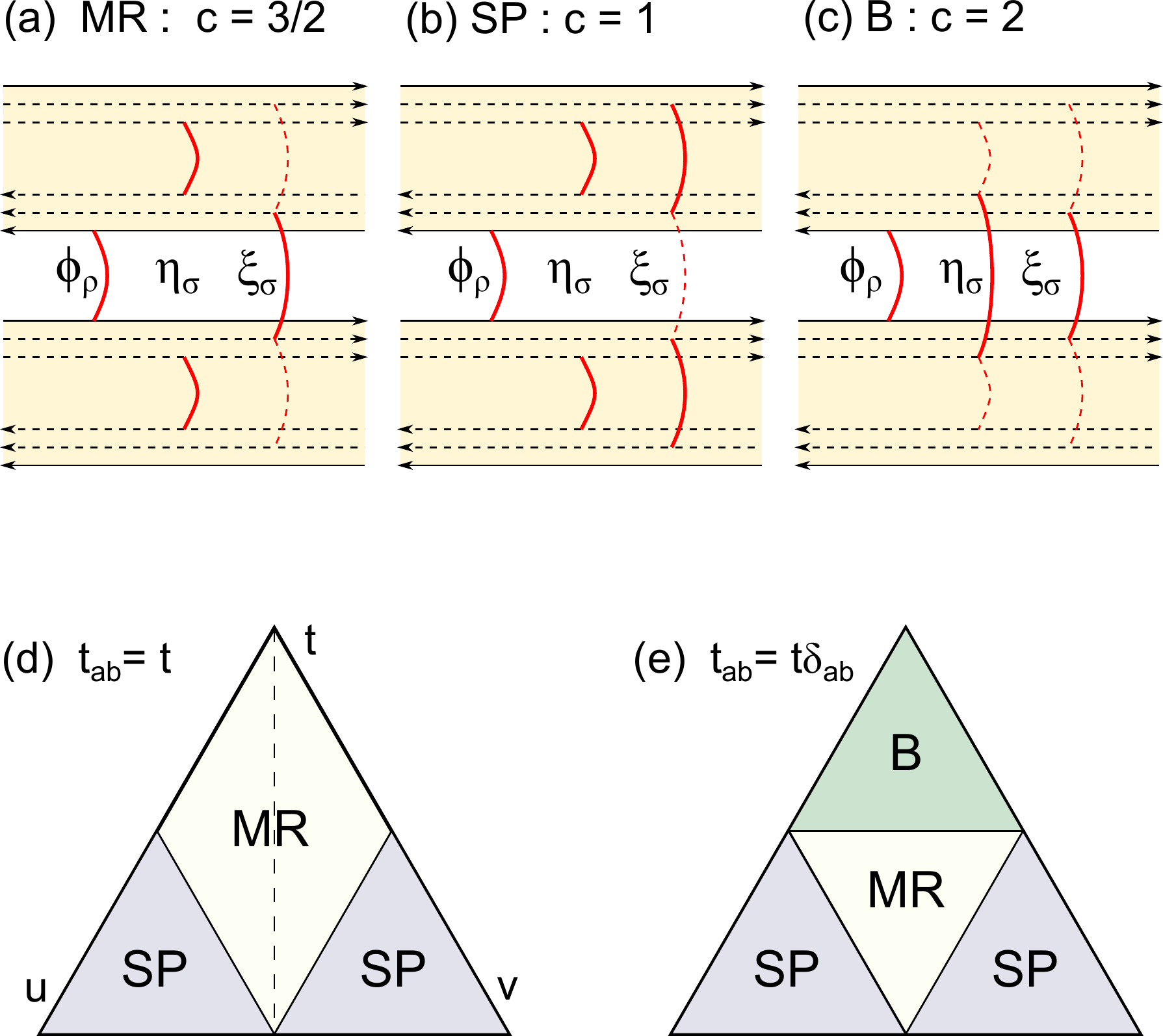}
\caption{The upper panels are schematic diagrams depicting the coupling of the chiral edge modes.  The solid lines represent the charge modes $\tilde\phi^{R/L}_{j,\rho}$.  The dashed lines represent the chiral Majorana modes $\xi^{R/L}_{j,\sigma}$ and $\eta^{R/L}_{j,\sigma}$.  Solid (dashed) arcs represent stronger (weaker) coupling.  (a) describes the Moore Read (MR) state, with c=3/2 edge states.  (b) describes a strong pairing (SP) state with $c=1$ edge states, and (c) describes a bilayer (B) state  with $c=2$ edge states. The lower panels are ternary phase diagrams as a function of $t$, $u$ and $v$ in the cases where (d) $t_{ab}=t$, independent of $a$ and $b$ and (e) $t_{ab}= t\delta_{ab}$.  The dashed line in (d) is the solvable line, where the decoupling of the chiral modes is perfect.}
\label{Fig 5}
\end{figure}

We now assume ${\cal H}_{\rm SLL}^0[\tilde\theta_{\ell,\rho},\tilde\varphi_{\ell,\rho}]$ is in a regime such that $\tilde t$ is relevant, and $\tilde\theta_{\ell,\rho}$ is pinned in a self consistent minimum of $\cos\tilde\theta_{\ell,\rho}$.  ${\cal H}$ then describes independent free fermion problems for $\xi_{j,\sigma}$ and $\eta_{j,\sigma}$.  The $\eta$ sector has a gap with a $k_y$ independent dispersion $E = \pm \sqrt{v^2 k_x^2 + (\tilde u+\tilde v)^2}$.  The
$\xi$ sector has dispersion $E = \pm \sqrt{v^2 k_x^2 + |\tilde t e^{i k_y} + \tilde u - \tilde v|^2}$ with a gap that closes at a point when $\tilde t =\pm |\tilde u - \tilde v|$ signaling a quantum phase transition.   The phase diagram is shown in Fig. \ref{Fig 5}d.

We identify the $\tilde t > |\tilde u  - \tilde v|$ phase with the Moore Read state\cite{mr,readgreen}.  Its physics is most transparent at the special point $\tilde u = \tilde v$ where the chiral Majorana modes $\eta^{R/L}_{j,\sigma}$ and $\xi^{R/L}_{j,\sigma}$ pair up with a pattern shown in Fig. \ref{Fig 5}a.  In this case, it is clear that it has a left over gapless chiral charge mode $\phi_{1,\rho}^R$ and a single chiral Majorana mode $\xi_{1,\sigma}^R$.  This is precisely the structure of the edge of the $\nu = 1$ bosonic Moore Read state\cite{fendley}.   Similar to the Abelian case, we may view the change of variables (\ref{chiral mr},\ref{chiral mr rhosigma},\ref{fermionize}) as transforming a SLL of coupled bosonic wires to a SLL of strips of $\nu=1$ Moore Read states coupled by their edge states.  In this case, however, the coupling of the edge states goes in ``both directions":  the $\eta^R_{j,\sigma}\eta^L_{j,\sigma}$ couples the edge states on a single strip, leaving behind the gapless edge states, while the $\cos\tilde\theta_{j+1/2,\rho} \xi^R_{j,\sigma} \xi^L_{j+1,\sigma}$ term couples the edge states on neighboring strips, leading to the bulk Moore Read state.

As in the case of the Abelian quantum Hall states, bulk quasiparticle excitations are associated with kinks in $\tilde\theta_{\ell,\rho}(x)$.  The present case is slightly different, though because the transformation $\varphi_{j,1} \rightarrow \varphi_{j,1}+2\pi$ (which connects equivalent states) translates to $\phi^{R/L}_{j,\rho/\sigma} \rightarrow \phi^{R/L}_{j,\rho/\sigma}+\pi$.  It then follows that the transformation $\tilde\theta_{j,\rho} \rightarrow \tilde\theta_{j,\rho}+\pi/2$, $(\xi_{j,\sigma},\eta_{j,\sigma}) \rightarrow -(\xi_{j,\sigma},\eta_{j,\sigma})$ connects equivalent states.  The elementary quasiparticle is thus associated with a kink in which $\tilde\theta_{\ell,\rho}(x)$ jumps by $\pi/2$, corresponding
to a charge $e/2$.  This introduces a domain wall where the mass term coupling $\xi^R_{j,\sigma}$ and $\xi^L_{j+1,\sigma}$ changes sign.  This binds a zero energy Majorana bound state, as is expected for the charge $e/2$ quasiparticles of the bosonic Moore Read state.   As in Section II.C.2, the quasiparticle tunneling operators can be related to the backscattering of bare electrons.  We defer the discussion of this to section IV.C.

The $\tilde t < |\tilde u  - \tilde v|$ phase corresponds to a strongly paired quantum Hall state of charge $2e$ bosons at filling $\nu=1/4$.  It is most easily understood in the limit $\tilde t << |\tilde u - \tilde v|$.  In this case, the Majorana modes pair up with the pattern in Fig. \ref{Fig 5}b, so that there are no gapless Majorana modes at the edge.  In this limit, individual bosons can not tunnel between wires because it excites the gapped $\xi_{j,\sigma}$ modes.  However, a pair of bosons can tunnel without disturbing the $\xi_{j,\sigma}$ sector.  The charge modes thus pair up leaving a single gapless chiral charge mode at the edge.

We now briefly discuss the case in which we relax our assumption about the equality of the different $t_{ab}$.  In this case, the $t_{ab}$ terms will have the general structure
\begin{equation}
\sum_{ab} t_{ab} {\cal O}^t_{j,ab} + h.c. =
i\left( \begin{array}{cc} \xi^R_{j,\sigma} & \eta^R_{j,\sigma} \end{array}\right) {\bf T}
\left( \begin{array}{c} \xi^L_{j+1,\sigma} \\ \eta^L_{j+1,\sigma} \end{array}\right),
\end{equation}
where the $2\times 2$ matrix $T_{nm} = \tilde t_{nm}\cos ( 2\tilde\theta_{\ell,\rho} + \beta_{ab})$ is characterized by a magnitude $\tilde t_{nm}$ and a phase $\beta_{nm}$, which depend on $t_{ab}$.  When $\tilde\theta_\rho$ is stiff, we may again analyze the problem by putting $\tilde\theta_\rho$ in a self consistent minimum and solving the non interacting fermion problem.

It is clear that due to the existence of a bulk energy gap, the phases discussed above will persist for a finite range in the more general parameter space.  However, another possible phase is possible where the neutral Majorana modes pair up in the manner shown in Fig \ref{Fig 5}c.  In this case the edge has a gapless charge mode and two gapless Majorana modes, or equivalently two gapless bosonic modes.

To see this, consider another special limit where $t_{ab} = t \delta_{ab}$, with $t$ real.  It then follows that $\tilde t_{nm} = \tilde t \delta_{nm} \cos 2\tilde\theta_{\ell,\rho}$, so that the $V$ has a term $\tilde t \cos 2\tilde\theta_{\ell,\rho} (\xi^R_{j,\sigma}\xi^L_{j+1,\sigma}+\eta^R_{j,\sigma}\eta^L_{j+1,\sigma})$.   The phase diagram for this case is shown in Fig \ref{Fig 5}e.  When $\tilde u = \tilde v$, the $\xi$ sector is gapped by $\tilde t$, while the $\eta$ sector involves competition between $\tilde t$ and $\tilde u+\tilde v$.  For $\tilde t < \tilde u+\tilde v$ we have the pairing in Fig. \ref{Fig 5}a, giving the Moore Read state, while for $\tilde t > \tilde u+ \tilde v$, we have the pairing in Fig. \ref{Fig 5}c, which has two bosonic edge modes.   This is most easily understood when $t_{ab} = t \delta_{ab}$ and $\tilde u=\tilde v = 0$, which simply corresponds to two decoupled $\nu=1/2$ bosonic quantum Hall states.

\subsection{Coset Construction}

The results of the previous section can be understood within the framework of the coset construction in conformal field theory\cite{coset,cft}.  This is useful because it helps us make contact with the work of Fradkin, Nayak and Shoutens\cite{fradkin} as well as subsequent generalizations.  It also introduces a framework that will allow us to generalize our construction to the Read Rezayi\cite{readrezayi} sequence of quantum Hall states.  Here we give a brief introduction to this well developed, but somewhat abstract, mathematical construction that emphasizes its physical meaning in the context of our coupled wire theory.

\begin{figure}
\includegraphics[width=3in]{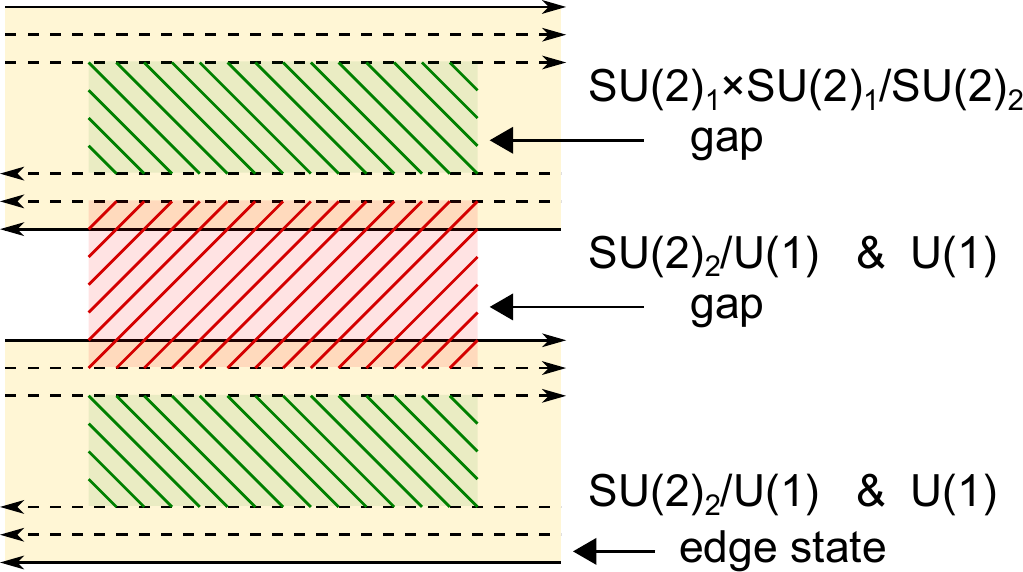}
\caption{Coupling of edge states at the decoupled point described by the dashed line in Fig. \ref{Fig 5}d.
The right and left moving $U(1)$ charge modes and the $SU(2)_2/U(1)$ Majorana fermion ($Z_2$ parafermion) modes are coupled on neighboring wires, while the right and left moving $SU(2)_1\times SU(2)_1/SU(2)_2$ modes are coupled on the same wire.  This pattern leaves $U(1)$ and $SU(2)_2/U(1)$ chiral modes at the edge.  This provides a concrete interpretation for the coset construction for the Moore Read state.}
\label{Fig 6}
\end{figure}

Our construction began with the SLL fixed point, which has bosonic modes $\tilde\phi^{R/L}_{j,a}$ on each wire.  The two right moving chiral modes on each wire define a conformal field theory with central charge $c=2$.  Through the sequence of transformations, the Hilbert space of these two modes was split into three pieces, described by $\tilde\phi^R_{j,\rho}$, $\xi_{j,\sigma}$ and $\eta_{j,\sigma}$.  The bosonic mode corresponds to $c=1$, while each of the Majorana modes has $c=1/2$.
At the point $u=v>0$ and $t_{ab}=t$, the decoupling is perfect.  The decomposition of the Hilbert space can be summarized by
\begin{equation}
2 = 1/2 + 1/2 + 1.
\label{2=1+1/2+1/2}
\end{equation}
The coupling terms we introduced allow the modes in the different sectors to pair up in different directions, as shown in Fig. \ref{Fig 6}.  This leads to a bulk gap, but leaves behind edge states in some of the sectors.  The Moore Read state thus has $c=3/2$ at the edge.

To understand this decomposition more generally, it is important to realize that at the
special filling factor $\nu=1/2$ the original chiral modes $\phi^R_a$ have an extra  symmetry because $\exp i\tilde\phi^R_a$ has scaling dimension $\Delta = 1$.  It follows that the operators
\begin{eqnarray}
J^R_{a,\pm} &=& e^{\pm i\tilde\phi^R_a}/(2\pi x_c), \\
J^R_{a,z} &=& \partial_x\tilde\phi^R_a/(2\pi).
\end{eqnarray}
generate an $SU(2)$ symmetry.  Each channel is thus described by a $SU(2)_1$ Wess Zumino Witten model.  The two channels together have
$SU(2)_1 \times SU(2)_1$ symmetry.  $SU(2)_1 \times SU(2)_1$ has a diagonal subgroup, $SU(2)_2$, generated by ${\bf J}^R = {\bf J}^R_1 + {\bf J}^R_2$.  In terms of the boson operators we have
\begin{eqnarray}
J^R_\pm &=& e^{\pm i\phi^R_{\rho}} \cos \phi^R_\sigma/(2\pi x_c), \\
J^R_z &=& \partial_x\phi^R_\rho/(2\pi).
\end{eqnarray}
$SU(2)_2$, in turn, has a subgroup $U(1)$ generated by $J^R_z$.

The coset construction allows a Wess Zumino Witten (WZW) model described by a group $G$ with subgroup $H$ to be divided into two pieces described by $G/H$ and $H$.  This means that the Hamiltonian can be written as the sum of two commuting terms, ${\cal H}_{G} = {\cal H}_{G/H} + {\cal H}_H$, so that the Hilbert space of eigenstates factorizes.  In the language of conformal field theory
the energy momentum tensor can be written $T_G = T_{G/H} + T_H$, and the components $T_{G/H}$ and $T_{H}$ have no singularities in their operator product expansion.  It follows that the central charge of a coset theory given simply by $c_{G/H} = c_G - c_H$.   Applied to the present problem, we have
\begin{eqnarray}
T_{SU(2)_1 \times SU(2)_1} = \hspace{2in}
\label{T=T+T} \\
T_{{SU(2)_1 \times SU(2)_1 / {SU(2)_2}}} + T_{{SU(2)_2 / U(1)}} + T_{U(1)}. \nonumber
\end{eqnarray}
Using the fact that $c_{SU(2)_k} = 3k/(k+2)$ it is simple to see that \eqref{T=T+T} is equivalent to \eqref{2=1+1/2+1/2}.

Consider the hopping term between wires, which for $t_{ab}=t$ can be written
\begin{equation}
t \sum_{ab} O^t_{j,ab} + h.c. = 8 t J^R_{+ j} J^L_{- j+1}.
\end{equation}
This term acts only in the $SU(2)_2$ sector for the pair of chiral modes, and leads to an energy gap in that sector.  Thus, if that were the only term (i.e. $u = v = 0$), then the $SU(2)_2/U(1)$ and $U(1)$ sectors would both be gapped, but each wire would retain $c=1/2$ chiral modes from the $SU(2)_1 \times SU(2)_1/SU(2)_2$ sector.

On the other hand, the term $(\tilde u + \tilde v) \sin\tilde\phi^R_{j,\sigma} \sin\tilde\phi^L_{j,\sigma}$ acts only in the $SU(2)_1 \times SU(2)_1/SU(2)_2$ sector.  Thus, if this is the only term (i.e. $t = u-v=0$), then the $SU(2)_1 \times SU(2)_1/SU(2)_2$ is gapped, while there will be $c=3/2$ chiral modes associated with the $SU(2)_2/U(1)$ and $U(1)$ sectors.  While this fact is clear from fermion representation \eqref{fermionize}, we will defer the proof that $\sin\phi^R_{j,\sigma}$ acts only in the $SU(2)_1 \times SU(2)_1/SU(2)_2$ sector to section IV.B, where it will be demonstrated in a more general context.

To summarize, the coset construction provides a way to fractionalize a $c=2$ Luttinger liquid into non trivial pieces.  When those pieces pair up in ``different directions", as in Fig. \ref{Fig 6}, the resulting fully gapped phase is a quantum Hall state with edge states that reflect the non trivial coset conformal field theory.  In the following section we will generalize this to develop a coupled wire construction for the Read Rezayi sequence of quantum Hall states, described by a $SU(2)_k$ theory.

Before doing that, though, we will conclude this section by generalizing the coupled wire construction of the bosonic Moore Read state to describe fermions.

\subsection{Generalization to fermions}

We now consider coupled wires of fermions.  Unfortunately, for a uniform magnetic field
there is no simple coupled wire construction for the $\nu = 1/2$ fermionic Moore Read state.
The tunneling terms that are allowed by momentum conservation either lead to a strongly paired Abelian quantum Hall state of charge $2e$ bosons, or they involve pairs of non commuting terms that can not be easily analyzed using the present methods.  Evidently, the Moore Read state is not sufficiently ``close" to the SLL fixed point for uniform field.

However, we found that if the magnetic field is staggered, so that the flux between neighboring wires alternates between two values, then a construction similar to the preceding section can be developed.   One can view this as a generalization of the two channel construction in the preceding section, where instead of having the two layers directly on top of one another, one layer is slid over relative to the other.  Equivalently, this can be viewed as a single layer system with
a staggered field as in Fig. \ref{Fig 7}a).  The system in Fig. \ref{Fig 4}a has a magnetic flux per unit length in units of the flux quantum $b=eaB/\hbar$ that alternates between $b$ and $0$.
Our more general construction then corresponds to sliding one layer relative to the other, so that the flux per length in units of the flux quantum  alternates between two values $b_1=\bar b + \delta b$ and $b_2 = \bar b - \delta b$.  The average flux $\bar b$ is related to the filling factor,  $\nu = 2 k_F /\bar b$.  The two channel bosonic problem then corresponds to $\delta b = \bar b$, while the uniform field corresponds to $\delta b=0 $.  We will show that when $\delta b = 2 k_F$ the allowed tunneling terms have a structure similar to that in (\ref{Ot}-\ref{Ov}).   This construction gives the $\nu=1/2$ fermionic Moore Read state for $\bar b = 4 k_F$, as well as the more general ``$q$-Pfaffian" state at $\nu=1/(1+q)$ for $\bar b = 2 k_F (1+q)$, where $q$ is even (odd) for bosons (fermions).   The state in this series with $q=-1$ is special, and corresponds to a $p+ip$ superconductor in zero net magnetic field.  In our construction, the modification of the state by changing the uniform component of the field $\bar b$ is reminiscent of modifying the Moore Read wavefunction by including a Jastrow factor that compensates the change in magnetic field\cite{mr}.

\subsubsection{$q$-pfaffian state}

\begin{figure}
\includegraphics[width=3in]{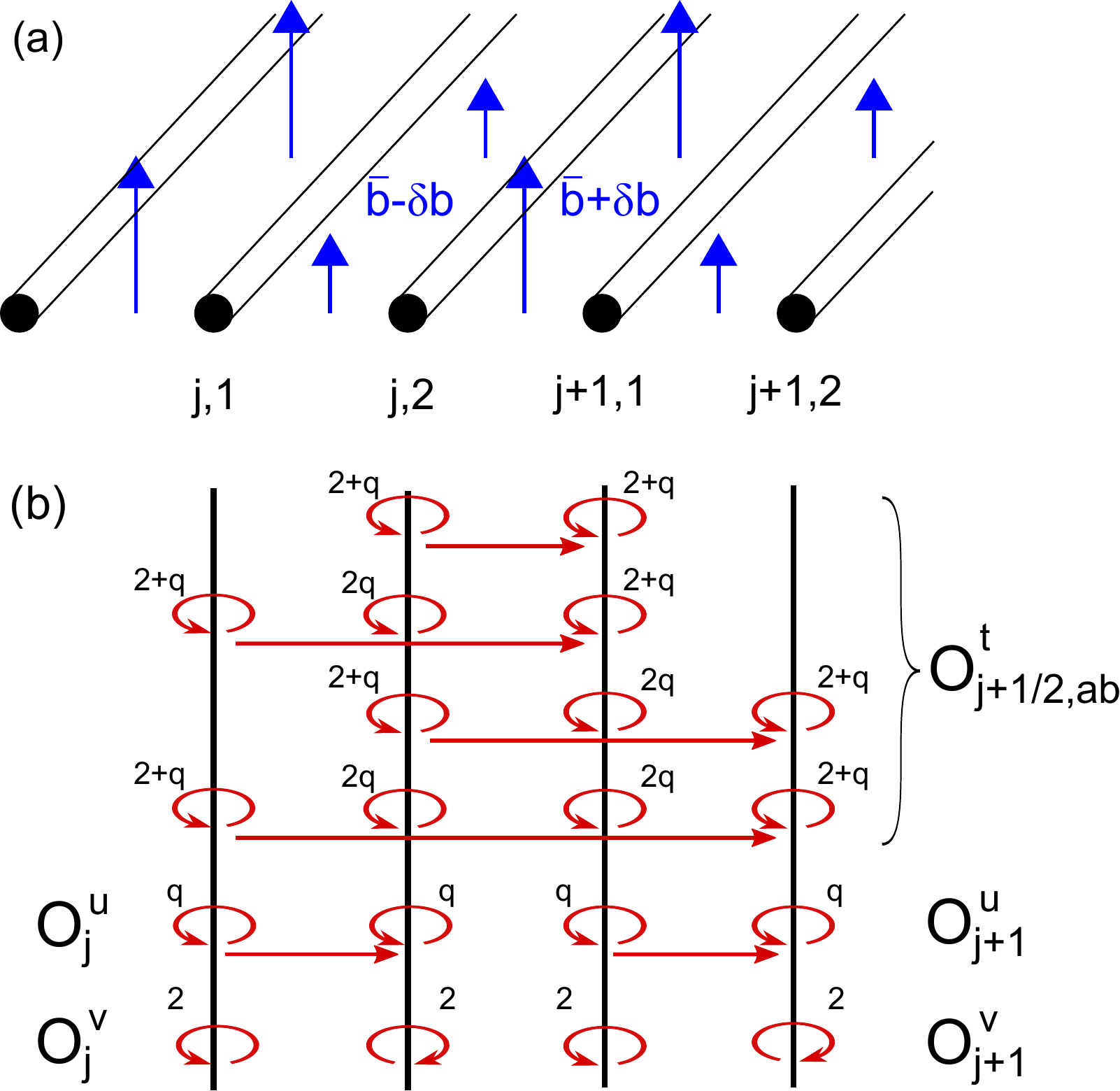}
\caption{(a) Coupled wire construction with a staggered magnetic field.  For $\delta b = \bar b$,
the theory is equivalent to the two component boson system shown in Fig \ref{Fig 4}a.  (b) Schematic diagram showing the coupling of the edge states, similar to Fig \ref{Fig 4}b.}
\label{Fig 7}
\end{figure}

Consider an array of wires with alternating magnetic flux, shown in Fig \ref{Fig 7}.   We parameterize the two fluxes as $2k_F (2+q)$ and $2 k_F q$.  We group the wires into pairs, indexed by $j$ and $a=1,2$.  The interaction terms then have a form similar to (\ref{Ot}-\ref{Ov}):
\begin{equation}
V = \sum_j \int dx\left( \sum_{ab=1}^2 t_{ab} {\cal O}^t_{j,ab}
+ u {\cal O}^u_j + v {\cal O}^v_j\right) + h.c.
\end{equation}
There are four terms coupling pair $j$ to $j+1$,
\begin{equation}
{\cal O}^t_{j,ab} = e^{i(\varphi_{j,a}-\varphi_{j+1,b}) + (q+2)(\theta_{j,a}+\theta_{j+1,b}))} Q_{j,ab}
\end{equation}
with
\begin{equation}
{\bf Q}_{j} = \left(\begin{array}{ll}
 e^{i 2q  \theta_{j,2}} & e^{i 2q  (\theta_{j,2}+ \theta_{j+1,1})} \\
1  & e^{i 2q  \theta_{j+1,1}}
 \end{array}\right).
\end{equation}
%\begin{eqnarray}
%Q_{j,11} &=& e^{i 2q  \theta_{j,2}} \\
%Q_{j,21} &=& 1 \\
%Q_{j,12} &=& e^{i 2q  (\theta_{j,2}+ \theta_{j+1,1})} \\
%Q_{j,22} &=& e^{i 2q  \theta_{j+1,1}}
%\end{eqnarray}
Two terms operate within a single pair.  The first involving tunneling an electron
between the two wires
\begin{equation}
{\cal O}^u_{j} = e^{i\left( \varphi_{j,1} - \varphi_{j,2} + q(\theta_{j,1}+\theta_{j,2})\right)}.
\end{equation}
The second giving an interaction between the $2k_F$ densities.
\begin{equation}
{\cal O}^v_{j} = e^{i\left( 2\theta_{j,1} - 2\theta_{j,2} \right)}.
\end{equation}
From \eqref{fermionmn} and \eqref{bosonmn}, it is clear that these interactions are appropriate for bosons (fermions) when $q$ is even (odd).

We can write the first term as
\begin{equation}
{\cal O}^t_{j,ab} = e^{i\tilde\phi^R_{j,a} - \tilde\phi_{j+1,b}^L}\\
\end{equation}
with
\begin{eqnarray}
\tilde\phi^R_{j,1} &=& \varphi_{j,1} + (q+2)\theta_{j,1} + 2q \theta_{j,2} \nonumber\\
\tilde\phi^L_{j,1} &=& \varphi_{j,1} - (q+2)\theta_{j,1} \nonumber \\
\tilde\phi^R_{j,2} &=& \varphi_{j,2} + (q+2)\theta_{j,2}  \label{tilde phirl q}\\
\tilde\phi^L_{j,2} &=& \varphi_{j,2} - (q+2)\theta_{j,2} - 2q \theta_{j,1} \nonumber
\end{eqnarray}
We next define the sum and difference variables
%\begin{eqnarray}
%\phi_{j,\rho}^R &=& \varphi_{j,1}+\varphi_{j,2}+(q+2)\theta_{j,1}+(3q+2)\theta_{j,2}  \\
%\phi_{j,\sigma}^R &=& \varphi_{j,1}-\varphi_{j,2}+(q+2)\theta_{j,1}+(q-2)\theta_{j,2} \\
%\phi_{j,\rho}^L &=& \varphi_{j,1}+\varphi_{j,2}-(3q+2)\theta_{j,1}-(q+2)\theta_{j,2} \\
%\phi_{j,\sigma}^L &=& \varphi_{j,1}-\varphi_{j,2}+(q-2)\theta_{j,1}+(q+2)\theta_{j,2}
%\end{eqnarray}
\begin{eqnarray}
\tilde\phi_{j,\rho}^{R/L} &=& (\tilde\phi_{j,1}^{R/L} + \tilde\phi_{j,2}^{R/L})/2,  \nonumber\\
\tilde\phi_{j,\sigma}^{R/L} &=& (\tilde\phi_{j,1}^{R/L} - \tilde\phi_{j,2}^{R/L})/2.
\label{phi rhosigma q}
\end{eqnarray}
These variables obey the commutation relations,
\begin{eqnarray}
\left[\partial_x\tilde\phi^p_{j,\rho}(x),\tilde\phi^{p'}_{j',\rho}(x')\right] &=& 2\pi i(1+q)  p \delta{pp'}\delta_{jj'}\delta_{xx'}
\label{phirho mr commutator}\\
\left[\partial_x\tilde\phi^p_{j,\sigma}(x),\tilde\phi^{p'}_{j',\sigma}(x')\right] &=& 2\pi i p \delta{pp'} \delta_{jj'}\delta_{xx'}
\label{phisigma mr commutator}
\end{eqnarray}
Note that the commutation relation for the $\sigma$ sector is identical to \eqref{phirl mr commutator}.  This allows us to proceed in the same manner as section III.A.  For the $\sigma$ sector to be unaltered, it was essential that the staggered field satisfy $\delta b = 2k_F$.  The charge sector, on the other hand is modified, and resembles that of the Laughlin state with $m=1+q$.

As in section III.A we focus on the case where $t_{ab}=t$, independent of $a$ and $b$.  Then
\begin{equation}
\sum_{ab} {\cal O}_{j,ab}^t = e^{i(\tilde\phi^R_{j,\rho}-\tilde\phi^L_{j+1,\rho})} \cos \tilde\phi_{j,\tilde\sigma}^R \cos\tilde\phi_{j+1,\sigma}^L,
\label{ot q}
\end{equation}
along with
\begin{eqnarray}
{\cal O}^u_j &=& e^{i(\tilde\phi^R_{j,\sigma} + \tilde\phi^L_{j,\sigma})}, \\
{\cal O}^v_j &=& e^{i(\tilde\phi^R_{j,\sigma} - \tilde\phi^L_{j,\sigma})}.
\end{eqnarray}

From this point the analysis proceeds exactly as in section III.A and will not be repeated.  The only difference is that the charge sector has a modified structure constant in \eqref{phirho mr commutator}.  This changes the exponents for tunneling electrons into the edge states, as well as the charge of the bulk quasiparticles.  As expected, for the $\nu=1/2$ Moore Read state ($q=1$), the quasiparticles have charge $e/4$.

\subsubsection{p+ip superconductor}

The special case $q=-1$ in the previous section corresponds to a fermion system with $\bar b = 0$.  This is not a quantum Hall state, but rather a $p+ip$ superconductor\cite{readgreen}, and the analysis is slightly different.
For $q=-1$, the transformation in \eqref{tilde phirl q} breaks down because $\tilde\phi^{R/L}_{j,1/2}$ become linearly dependent.  This can be seen from the fact that $\tilde\phi^R_{j,\rho}$ and $\tilde\phi^L_{j,\rho}$, defined in \eqref{phi rhosigma q} are in fact the same operator.

We proceed by defining $\tilde\phi^R_{j,\sigma}$ and $\tilde\phi^L_{j,\sigma}$ as in \eqref{tilde phirl q}, but replacing $\tilde\phi^R_{j,\rho}$ and $\tilde\phi^L_{j,\rho}$ by
\begin{eqnarray}
\tilde\varphi_{j,\rho} &=& \left(\varphi_{j,1} + \varphi_{j,2} + q(\theta_{j,2}-\theta_{j,1})\right)/2 \nonumber\\
\tilde\theta_{j,\rho} &=& \theta_{j,1} + \theta_{j,2}.
\end{eqnarray}
These satisfy $[\partial_x\tilde\theta_{j,\rho}(x),\tilde\varphi_{j',\rho}(x')] = i\pi \delta_{jj'}\delta_{xx'}$.
It follows that $\tilde\phi^{R/L}_{j,\rho} = \tilde\varphi_{j,\rho} \pm (1+q) \tilde\theta_{j,\rho}$.
For $q=-1$ the analysis is then the same, except that \eqref{ot q} becomes
\begin{equation}
\sum_{ab} O^t_{j,ab} = e^{\tilde\varphi_{j,\rho} - \tilde\varphi_{j+1,\rho}} \cos\tilde\phi_{j,\sigma}^R \cos\tilde\phi_{j+1,\sigma}^L.
\end{equation}

This term has precisely the form of the tunneling of electrons between the edge states of strips of $p+ip$ superconductor.  When $O^t_{j,ab}$ flows to strong coupling the a Josephson coupling $\cos 2(\tilde\phi_{j,\sigma}-\tilde\phi_{j+1,\sigma})$ will be generated, and the phases $\varphi_{j,\rho}$ on neighboring will lock together.  However, unlike the quantum Hall case, there will be a gapless bulk collective mode associated with slowly varying fluctuations in $\varphi_{j,\rho}$.  The neutral sector is identical to that of the Moore Read state, however, for the charge mode, the gapless edge mode in the quantum Hall case is replaced by a gapless bulk collective mode.

\section{Read Rezayi Sequence}

In this section, we will generalize the coupled wire construction to describe the Read Rezayi sequence of states\cite{readrezayi}.  This sequence includes the Moore Read state for $k=2$, as well as other states, which are described in terms of the $Z_k$ parafermion conformal field theory\cite{zamolodchikov}.  As in the previous section, the analysis is simplest for bosons.  Following the analysis of Fradkin, Nayak and Shoutens\cite{fradkin}, we thus consider $k$ channels of bosons, which are each at filling $\nu=1/2$, so that the total filling factor is $k/2$.  At the end of this section we will briefly describe the generalization, similar to section III.C, which gives the known Read Rezayi states at $\nu = k/(2 + k q)$, where $q$ is even (odd) for bosons (fermions).

\subsection{Bosons at $\nu = k/2$}

Consider coupled wires of $k$ channel bosons.  The analysis is similar to Section III.A, except now each wire is characterized by $\theta_{j,a}$ and $\varphi_{j,a}$ satisfying \eqref{thetaphi mr commutator} for $a=1, ..., k$.    The Hamiltonian is again ${\cal H}^0_{\rm SLL}(\theta,\phi) + V$ with
\begin{eqnarray}
V = \sum_j \int dx\left( \sum_{ab=1}^k t_{ab} {\cal O}^t_{j,ab}\right.
+ \hspace{.8in}\nonumber\\
\left.\sum_{a<b=1}^k u_{ab} {\cal O}^u_{j,ab} + v_{ab} {\cal O}^v_{j,ab}\right) + h.c.
\label{v rr}
\end{eqnarray}
The interaction coupling neighboring wires
\begin{equation}
{\cal O}^t_{j;ab} = e^{i\left (\varphi_{j,a} - \varphi_{j+1,b} + 2(\theta_{j,a}+\theta_{j+1,b})\right)}
\label{Ot rr}
\end{equation}
is the same as before, while the interactions operating within a single wire
come in more varieties,
\begin{equation}
{\cal O}^u_{j,ab} = e^{i\left( \varphi_{j,a} - \varphi_{j,b} \right)},
\label{Ou rr}
\end{equation}
as well as an interaction that locks the ``$2k_F$" densities of the two channels,
\begin{equation}
{\cal O}^v_{j,ab} = e^{i\left( 2\theta_{j,a} - 2\theta_{j,b} \right)}.
\label{Ov rr}
\end{equation}

As in section III.A we first define chiral boson modes
\begin{eqnarray}
\tilde\phi^R_{j,a} = (\varphi_{j,a} + 2\theta_{j,a})/\sqrt{2} \nonumber\\
\tilde\phi^L_{j,a} = (\varphi_{j,a} - 2\theta_{j,a})/\sqrt{2} .
\end{eqnarray}
For later convenience, this definition differs by a factor of $\sqrt{2}$ from
the modes defined in Eq. \ref{chiral mr}.
We then introduce a charge mode $\tilde\phi^{R/L}_{j,\rho}$ and $k-1$ neutral modes
$\vec\phi^{R/L}_{j,\sigma}$ by writing
\begin{equation}
\tilde\phi^{R/L}_{j,\mu} = \left( \begin{array}{c} \tilde\phi^{R/L}_{j,\rho}\\ \vec\phi^{R/L}_{j,\sigma} \end{array}\right)_\mu,
\end{equation}
with
\begin{equation}
\tilde\phi^{R/L}_{j,\mu} =  \sum_{a=1}^k O_{\mu a} \tilde\phi^{R/L}_{j,a}.
\end{equation}
$O_{\mu a}$ is an orthogonal matrix that has the form
\begin{equation}
O_{\mu a} = \left(  \begin{array}{c}  1/\sqrt{k} \\ \vec d_a \end{array}\right)_\mu
\end{equation}
where $\vec d_a$ are a set of $k$ vectors with $k-1$ components that satisfy
\begin{eqnarray}
\sum_a \vec d_a &=& 0
\label{d sum}\\
\sum_a d_a^\alpha d_a^\beta &=& \delta_{\alpha\beta}\\
\vec d_a \cdot \vec d_b &=& \delta_{ab} - 1/k.
\label{d prod}
\end{eqnarray}
$\vec d_a$ may be viewed as the unit vector in the $a$ direction projected into the plane perpendicular to $(1,1,...,1)$.  For example, for $k=3$ they form $3$ planar vectors oriented at $120^\circ$,
\begin{eqnarray}
\vec d_1 = \left(\begin{array}{c} 1\over\sqrt{2}\\1\over\sqrt{6} \end{array}\right), \
\vec d_2 = \left(\begin{array}{c} -1\over\sqrt{2}\\1\over\sqrt{6} \end{array}\right), \
\vec d_3 = \left(\begin{array}{c} 0 \\-2\over\sqrt{6} \end{array}\right).
\end{eqnarray}

The transformation then has the explicit form,
\begin{eqnarray}
\tilde\phi^{R/L}_{j,\rho} &=& {1\over \sqrt{k}}\sum_{a=1}^k \tilde\phi^{R/L}_{j,a} \nonumber\\
\vec\phi^{R/L}_{j,\sigma} &=& \sum_{a=1}^k \vec d_a  \tilde\phi^{R/L}_{j,a}.
\label{phirl rhosigma rr}
\end{eqnarray}
along with
\begin{equation}
\tilde\phi^{R/L}_{j,a} = \left({1 \over\sqrt{k}}\tilde\phi^{R/L}_{j,\rho} + \vec d_a \cdot \vec\phi^{R/L}_{j,\sigma}\right).
\end{equation}
For $k=2$ the charge and spin modes $\tilde\phi^{R/L}_{j,\mu}$ are identical to the corresponding modes defined in Eq. \ref{chiral mr rhosigma} in Section III.A (though $\tilde\phi^{R/L}_{j,a}$ differ by $\sqrt{2}$.   The charge and neutral modes satisfy
\begin{equation}
\left[\partial_x\tilde\phi_{j,\mu}^p(x),\tilde\phi_{j',\mu'}^{p'}(x')\right] = 2\pi i p\delta_{pp'}\delta_{\mu\mu'} \delta_{jj'} \delta_{xx'}
\end{equation}
%Note that for $k=2$ the orthogonal rotation is slightly different from the transformation %().  In the present section $\phi^{R/L}_{j,\rho}$ and $\phi^{R/L}_{j,\sigma}$ differ from %their counterparts in section () by a factor of $\sqrt{2}$, which leads to the factor of %two difference between the commutators in () and ().

Expressed in these variables, the interaction terms have the form,
\begin{equation}
{\cal O}^t_{j,ab} = e^{i\sqrt{2/k}(\tilde\phi^R_{j,\rho}-\tilde\phi^L_{j+1,\rho})} e^{i \sqrt{2}(\vec d_a \cdot \vec\phi^R_{j,\sigma} - \vec d_b \cdot \vec\phi^L_{j+1,\sigma})}
\end{equation}
\begin{eqnarray}
{\cal O}^u_{j,ab} &=& e^{i (\vec d_a - \vec d_b)\cdot (\vec\phi^R_{j,\sigma} + \vec\phi^L_{j,\sigma})/\sqrt{2}}
\label{ou d rr}\\
{\cal O}^v_{j,ab} &=& e^{i (\vec d_a - \vec d_b)\cdot (\vec\phi^R_{j,\sigma} - \vec\phi^L_{j,\sigma})/\sqrt{2}}.
\label{ov d rr}
\end{eqnarray}

We will now focus of the special case in which $t_{ab}=t$ are independent of $a$ and $b$.  Then,
\begin{eqnarray}
V = \sum_j \int dx t e^{i\sqrt{2/k}(\tilde\phi^R_{j,\rho}-\tilde\phi^L_{j+1,\rho})} \Psi^R_{j} \Psi^L_{j+1} + \nonumber\\
\sum_{ab} (u_{ab}+v_{ab}) i\Upsilon^R_{j, ab} \Upsilon^L_{j, ab} + (u_{ab}-v_{ab}) i\Xi^R_{j, ab} \Xi^L_{j, ab}
\end{eqnarray}
where
\begin{equation}
\Psi^{R/L}_{j} = \sum_a \exp \left[i \sqrt{2}\vec d_a \cdot \vec\phi^{R/L}_{j,\sigma}\right]
\label{Psi}
\end{equation}
and
\begin{eqnarray}
\Upsilon^{R/L}_{j, ab} &=&  \sin \left[{1\over \sqrt{2}}(\vec d_a - \vec d_b) \cdot \vec\phi^{R/L}_{j,\sigma}\right]
\label{upsilon}\\
\Xi^{R/L}_{j, ab} &=&  \cos \left[{1\over \sqrt{2}}(\vec d_a - \vec d_b) \cdot \vec\phi^{R/L}_{j,\sigma}\right].
\label{xi}
\end{eqnarray}

In the following we will show that at the special point $u=v$ a decoupling similar to what occurred in Section III.B occurs.  To establish this, we will first use the coset construction to show how the $k$ chiral modes on each wire decouple into separate sectors.  We will then show that $\Psi^{R/L}_j$ acts only in one sector, while $\Upsilon^{R/L}_j$ acts only in the other.  The coupling terms in \eqref{v rr} then lead to gaps in which the different sectors are paired in different directions, leaving behind non trivial edge states.  $\Psi^{R/L}_j$ will be identified as a $Z_k$ parafermion operator.  The coupling term $\Upsilon^{R}_{j,ab}\Upsilon^{L}_{j,ab}$, on the other hand, leads to a theory on an individual wire which can be identified with the critical point of a $Z_k$ model, which is a particular $k$ state generalization of the Ising and 3 state Potts model.

\subsection{Coset Construction and Primary Fields}

Each wire is characterized by $k$ right and left moving chiral modes, which individually are described by a $SU(2)_1$ WZW model.  As in the previous section, $[SU(2)_1]^k$ can be decoupled by considering the diagonal sub algebra $SU(2)_k$.  This leads to the following decomposition of the energy momentum tensor
\begin{eqnarray}
T_{[SU(2)_1]^k} = \hspace{2in}
\label{T level k}\\
T_{{[SU(2)_1]^k / {SU(2)_k}}} + T_{{SU(2)_k / U(1)}} + T_{U(1)}. \nonumber
\end{eqnarray}
In terms of the central charge, this is equivalent to
\begin{equation}
k = {k(k-1)\over{k+2}} + {2(k-1)\over{k+2}} + 1
\end{equation}
Clearly $k=2$ reduces to \eqref{2=1+1/2+1/2}.  For $k=3$, we have $3=6/5+4/5+1$.
The $SU(2)_k/U(1)$ sector is precisely the $Z_k$ parafermion theory introduced by Zamolodchikov and Fateev\cite{zamolodchikov}.  The $c=k$ theory thus decomposes into a $U(1)$ charge sector, a
$SU(2)_k/U(1)$ parafermion sector and a third sector described by the coset $[SU(2)]^k/SU(2)_k$.

\begin{figure}
\includegraphics[width=3in]{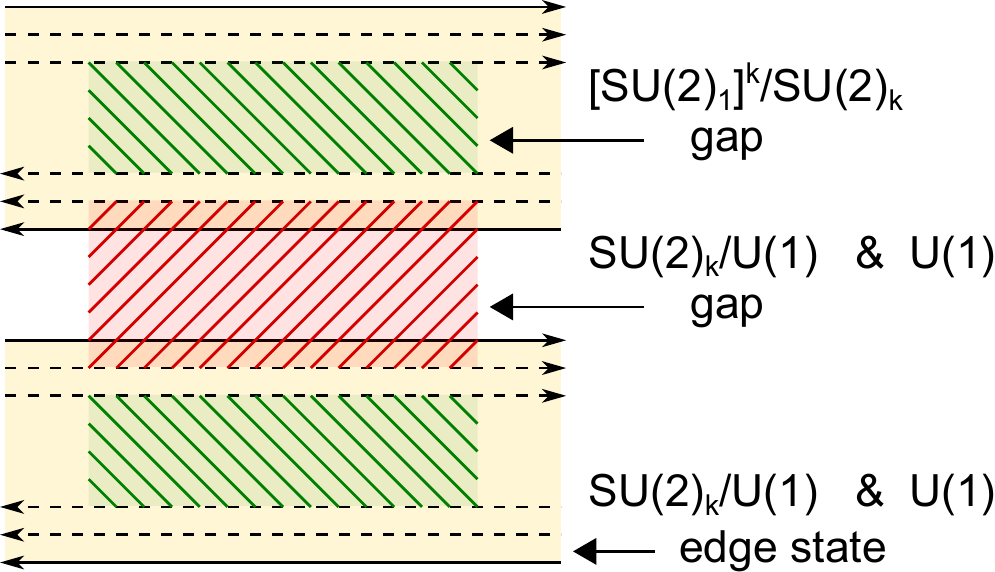}
\caption{Along the solvable line $u_{ab}=v_{ab}$ the right and left moving $U(1)$ charge modes and the $SU(2)_k/U(1)$ $Z_k$ parafermion modes are coupled only on neighboring wires, while the right and left moving $[SU(2)_1]^k/SU(2)_k$ modes are coupled only on the same wire.  This pattern leaves $U(1)$ and $SU(2)_k/U(1)$ chiral modes at the edge.  This provides a concrete interpretation for the coset construction for the Read Rezayi state.}
\label{Fig 8}
\end{figure}

To show this decoupling explicitly, and to demonstrate that $\Psi$ and $\Upsilon$ are primary fields that act only in the decoupled sectors, we explicitly construct the energy momentum tensors.
Here we consider the right moving chiral sector on a single wire $j$ and omit the superscript $R$ and subscript $j$.  The chiral channel is described by the Hamiltonian
\begin{equation}
{\cal H} = \int dx {v_F\over 4\pi}\left( (\partial_z\tilde\phi_\rho)^2 + (\partial_z\vec\phi_\sigma)^2 \right).
\label{h density rr}
\end{equation}
In order make contact with the conformal field theory literature, we focus on the energy momentum tensor, which is closely related to the Hamiltonian density,
\begin{equation}
T_{[SU(2)_1]^k} = -{1\over 2}\left( (\partial_z\tilde\phi_\rho)^2 + (\partial_z\vec\phi_\sigma)^2 \right).
\label{T [SU(2)_1]^k}
\end{equation}
The translation between the Hamiltonian formulation and the conformal field theory is briefly reviewed in Appendix B, where we also show that $T_{[SU(2)_1]^k}$ decomposes into three terms, according to \eqref{T level k}, with
\begin{equation}
T_{U(1)} = -{1\over 2} (\partial_z\phi_\rho)^2,
\label{tu1}
\end{equation}
\begin{eqnarray}
T_{SU(2)_k/U(1)} = \hspace{1.8in}\nonumber\\
{1\over{k+2}}\left( -(\partial_z\vec\phi_\sigma)^2 + \sum_{a\ne b} e^{i\sqrt{2}(\vec d_a-\vec d_b)\cdot\vec\phi_\sigma}\right)
\label{tparafermion}
\end{eqnarray}
and
\begin{eqnarray}
T_{[SU(2)_1]^k/SU(2)_k} = \hspace{1.5in}\nonumber\\
{1\over{k+2}}\left(-{k\over 2} (\partial_z\vec\phi_\sigma)^2 - \sum_{a \ne b} e^{i\sqrt{2}(\vec d_a-\vec d_b)\cdot\vec\phi_\sigma}\right).
\label{tcoset}
\end{eqnarray}
The non trivial content of this decoupling is that the three energy momentum tensors have no singular terms in their operator product expansion.  In a Hamiltonian formalism, this means that the Hamiltonian \eqref{h density rr}, splits into three commuting pieces, ${\cal H} = {\cal H}_{U(1)}+{\cal H}_{SU(2)_k/U(1)} + {\cal H}_{[SU(2)_1]^k/SU(2)_k}$.  This decoupling has also appeared in a somewhat different context in Ref. \onlinecite{affleck}

We now show that $\Psi$ and $\Upsilon$ act only in a single sector.  This is done by computing the operator product expansion with the $T$'s.  Details of the calculation are in Appendix C.  We find that for $z \rightarrow w$ the singular terms in the operator product expansion are
\begin{eqnarray}
T_{SU(2)_k/U(1)}(z) \Psi(w) = \hspace{1in} \nonumber\\
{1-1/k\over{(z-w)^2}} \Psi(z) + {1\over{(z-w)}} \partial_z\Psi(z),
\label{ope Psi1}
\end{eqnarray}

\begin{equation}
T_{U(1)}(z)\Psi(w) = T_{[SU(2)_1]^k/SU(2)_2} \Psi(w) = 0.
\label{ope Psi2}
\end{equation}
Eqs. \ref{ope Psi1} and \ref{ope Psi2} shows that $\Psi$ is the primary field of the $SU(2)_k/U(1)$ theory with scaling dimension $1-1/k$, known as a $Z_k$ parafermion operator.  $\Psi$ is a generalization of the Majorana fermion, which can be regarded as a $Z_2$ parafermion.  The fact that there no singular terms in the OPE for the other two sectors means that $\Psi$ acts only in the $SU(2)_k/U(1)$ parafermion sector.  In a Hamiltonian formulation, we would have $[\Psi,{\cal H}_{U(1)}]=[\Psi,{\cal H}_{[SU(2)_1]^k/SU(2)_k}]=0$.  A mass term $\Psi^{R\dagger}\Psi^L$ has scaling dimension $2-2/k$, and is relevant.  It leads to an energy gap in the parafermion sector.

For $\Upsilon_{ab}$ we find
\begin{eqnarray}
T_{[SU(2)_1]^k/SU(2)_k}(z) \Upsilon_{ab}(w) = \hspace{1in}
\label{ope Upsilon 1}\\
{1/2\over{(z-w)^2}} \Upsilon_{ab}(z) + {1\over{(z-w)}} \partial_z\Upsilon_{ab}(z), \nonumber
\end{eqnarray}
\begin{equation}
T_{U(1)}(z)\Upsilon_{ab}(w) = T_{SU(2)_k/U(1)}(z)\Upsilon_{ab}(w) = 0.
\label{ope Upsilon 2}
\end{equation}
This shows that $\Upsilon_{ab}$ are primary fields of the $[SU(2)]^k/SU(2)_k$ sector with scaling dimension $1/2$, and do not act in the $SU(2)_k/U(1)$ or the $U(1)$ sectors.   A mass term $\Upsilon^R \Upsilon^L$ has dimension $1$ and leads to an energy gap in the $[SU(2)]^k/SU(2)_k$ sector.

We have also computed the OPE's for $\Xi_{ab}$, defined in \eqref{xi}.  Unlike $\Psi$ and $\Upsilon$, though, $\Xi$ is not primary and acts non trivially in both the $[SU(2)]^k/SU(2)_k$ and the $SU(2)_k/U(1)$ sectors.  Thus, unlike the $k=2$ case, it is not clear how $\Xi^R\Xi^L$ competes with the other terms.   Nonetheless, on the special line $u_{ab}=v_{ab}$, $t_{ab}=t$ the $\Xi_{ab}$ term is absent, and we have the decoupling shown in Fig. \ref{Fig 9}, in which the $U(1)$ and $SU(2)_k/U(1)$ sectors are gapped across wires, while the $[SU(2)]^k/SU(2)_k$ sector is gapped within a wire.  This gives the Read Rezayi state, which has a left over gapless edge state with a
$U(1)$ charge mode and a $SU(2)_k/U(1)$ $Z_k$ parafermion mode.

%In section (), we will argue that when the $[SU(2)]^k/SU(2)_k$ is gapped by $\Upsilon^R %\Upsilon^L$, the operator $\Xi_{ab}$ projected into the remaining gapless $SU(2)_k$ sector becomes %a primary field of $SU(2)_k/U(1)$, which corresponds to the energy operator $\epsilon$.

\subsection{Quasiparticle Operators}

To construct quasiparticle operators we follow the logic of Section II.C.2 and consider the $2k_F$ backscattering of bare particles on channel $a$ of wire $j$,
\begin{eqnarray}
\chi_{j,a} &=& e^{2i\theta_{j,a}} \\
&=& e^{\sqrt{2/k}(\tilde\phi^R_{j,\rho}-\tilde\phi^L_{j,\rho}) + \sqrt{2} \vec d_a\cdot(\vec\phi^R_{j,\sigma}-\vec\phi^L_{j,\sigma})}
\end{eqnarray}
We thus define
\begin{equation}
\Psi^{R/L}_{QP,j+1/2,a} = e^{i\sqrt{2/k}\tilde\phi^{R/L}_{j/j+1,\rho}}\Sigma^{R/L}_{j/j+1,a}
\end{equation}
where
\begin{equation}
\Sigma^{R/L}_{j,a} = \exp\left[ i\sqrt{2}\vec d_a\cdot\vec\phi^{R/L}_{j,\sigma}\right].
\end{equation}
$\Psi^{R/L}_{QP,j+1/2,a}$ creates a quasiparticle with charge $1/2$ (check!) with a non trivial action in the neutral sector.

Like $\Xi_{ab}$, the operators $\Sigma_{a}$ are not primary, and acts in both the $[SU(2)]^k/SU(2)_k$ and $SU(2)_k/U(1)$ sectors.  Nonetheless, in the next section we will argue that when the $[SU(2)]^k/SU(2)_k$ sector is gapped, $\Sigma_a$ acts as a primary field projected into the parafermion sector, which corresponds to the spin operator $\sigma$.

\subsection{Relation to $Z_k$ Statistical Mechanics Model}

On a single wire, the mass term $(u_{ab}+v_{ab})\Upsilon_{ab}^R\Upsilon_{ab}^L$ opens a gap and leaves behind a $SU(2)_k$ gapless edge state with a charge mode and a $Z_k$ parafermion.  The $Z_k$ parafermion conformal field theory is known to describe the critical point of a $Z_k$ generalization of the Ising model\cite{zamolodchikov}.  For $k=3$ it is the 3 state Potts model. Similarly, for $k>3$ it is a particular version of a $Z_k$ symmetric $k$ state generalized Potts model.  In this section we show that our bosonized representation provides a simple and intuitively appealing way to understand this connection.  This allows us to identify the projected operators $\Sigma_a$ with the primary fields  $\sigma$ of the $Z_k$ parafermion model.

We begin by rewriting the mass term by introducing new variables for the single wire,
\begin{eqnarray}
\vec\varphi_\sigma &=& (\vec\phi_\sigma^R + \vec\phi_\sigma^L)/2, \nonumber \\
\vec\theta_\sigma &=& (\vec\phi_\sigma^R - \vec\phi_\sigma^L)/2.
\end{eqnarray}
These variables satisfy
\begin{equation}
[ \partial_x\theta_\sigma^\alpha(x), \varphi_\sigma^\beta(x')] = i\pi \delta_{\alpha\beta} \delta_{xx'}.
\label{thetaphi commutator zk}
\end{equation}
The Hamiltonian for the neutral sector of a single wire then has the form
\begin{eqnarray}
{\cal H} = {v\over{2\pi}}\left( (\partial_x\vec\theta_\sigma)^2 + (\partial_x\vec\varphi_\sigma)^2\right) + \hspace{1in}\nonumber \\
\sum_{ab} u_{ab} \cos \sqrt{2}\vec d_{ab}\cdot\vec\theta_\sigma + v_{ab}\cos \sqrt{2}\vec d_{ab}\cdot \vec \varphi_\sigma .
\end{eqnarray}
where we have abbreviated $\vec d_{ab} = \vec d_a - \vec d_b$.

First consider the simplest case $k=2$, where $\vec\theta_\sigma$ and $\vec\varphi_\sigma$ have a single component, and $d_{12} -d_{21} = \sqrt{2}$.  Then we have
\begin{equation}
{\cal H} = {v\over{2\pi}}\left( (\partial_x\theta_\sigma)^2 + (\partial_x\varphi_\sigma)^2\right)
- u \cos 2\theta_\sigma -  v \cos 2\varphi_\sigma
\label{hthetaphi zk}
\end{equation}
Viewed as a transfer matrix for the partition function of an anisotropic statistical mechanics problem, this Hamiltonian gives a well known representation of the 2D Ising model\cite{jose}.  This can be understood by first considering $u=0$.  This describes the 2D XY model with order parameter $(\cos\theta_\sigma,\sin\theta_\sigma)$.   For $v=0$, $\theta_\sigma$ is a non compact variable, so there are no vortices.  From \eqref{thetaphi commutator zk} it can be seen that $\exp\pm 2i\varphi_\sigma(x,\tau)$ creates a vortex where $\theta_\sigma$ winds by $\pm 2\pi$ around $(x,\tau)$.  $v$ is thus the fugacity for vortices, and its presence makes $\theta_\sigma$ an angular variable defined modulo $2\pi$.  Integrating out $\theta_\sigma$ gives the sine gordon representation of the XY model.  For nonzero $u$, $-\cos 2\theta_\sigma$ introduces an Ising anisotropy into the XY model.  For large $u$ $\theta_\sigma$ is pinned in the minima of this potential at $\theta = n\pi$.  Due to the presence of $v$, only two of those minima are distinct.   For $u\ne v$, since both $u$ and $v$ are relevant, the system flows at low energy to a strong coupling phase in which either $\theta_\sigma$ or $\phi_\sigma$ is pinned.  The symmetry under $u\leftrightarrow v$ and $\theta_\sigma \leftrightarrow \phi_\sigma$ is precisely the Kramers Wannier duality of the Ising model.  At the self dual point $u=v$, the system at low energy flows to the fixed point of the Ising critical point.

\begin{figure}
\includegraphics[width=3in]{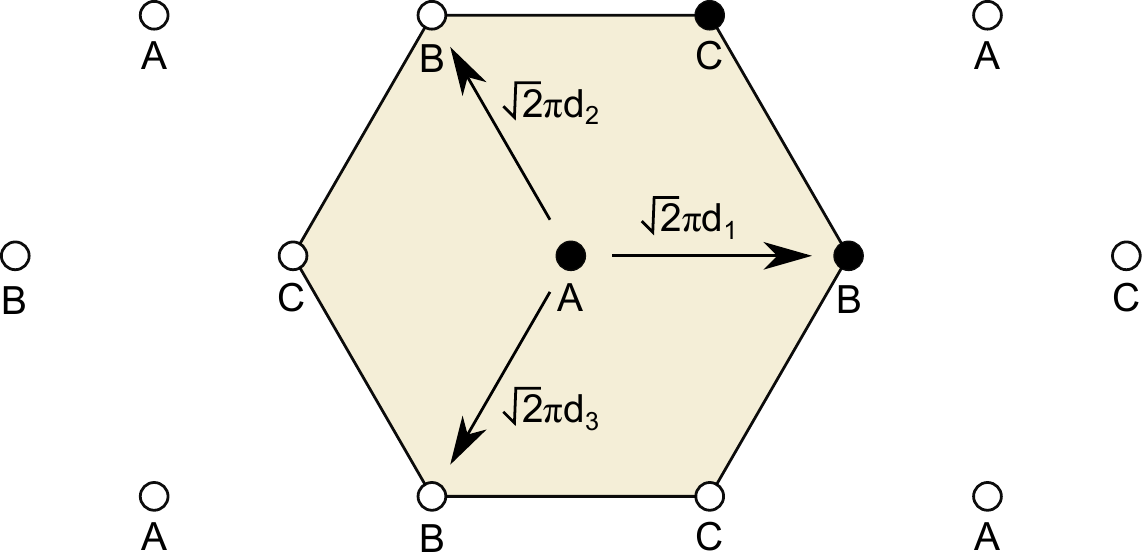}
\caption{Lattice of minima of the periodic potential $-\sum_{ab} \cos\sqrt{2} \vec d_{ab}\cdot\vec\theta_\sigma$ as a function of $\vec\theta_\sigma$.  The shaded region shows the set of distinct values of $\vec\theta_\sigma$, which is compactified to a torus due to the vortices created by $\exp i\sqrt{2}\vec d_{ab}\cdot\vec\varphi_\sigma$.  There are thus three distinct, but equivalent minima, labeled A, B and C.}
\label{Fig 9}
\end{figure}

For $k>2$, a similar interpretation is possible.  Now, however, $\vec\theta_\sigma$ lives in $k-1$ dimensions.  $\exp i \sqrt{2} \vec d_{ab}\cdot\vec\varphi_{\sigma}$ creates vortices around which $\vec\theta \rightarrow \vec\theta_\sigma + \sqrt{2}\pi \vec d_{ab}$.  This compactifies $\vec\theta_\sigma$, so that it is defined on a $k-1$ dimensional torus.    $\cos \sqrt{2} \vec d_{ab}\cdot\vec\theta_\sigma$ introduces a periodic potential for $\vec\theta_\sigma$, and pins $\vec\theta$ in its minima.

For $k=3$ the minima of the periodic potential are shown in Fig. \ref{Fig 9}.  They form a triangular lattice with lattice constant $2\pi/\sqrt{3}$.  The compactification of $\vec\theta_\sigma$ is associated with a larger triangular lattice with lattice constant $2\pi$.  This identifies points on the original lattice that are on the same $\sqrt{3}$ sublattice.  There are $3$ distinct minima.  We thus have a two dimensional generalization of the XY model (where the order parameter is defined on a torus $T^2$), with a 3 state anisotropy.  Since both the vortices and the anisotropy are relevant (the periodic potential has scaling dimension $1$), this leads to a 3 state generalization of the Ising model with $Z_3$ symmetry, which is uniquely specified by the 3 state Potts model.
Again, the critical point appears at the self dual point $u_{ab}=v_{ab}$.

For $k=4$, the minima of the periodic potential form a three dimensional fcc lattice.  The fcc lattice can be viewed as a larger bcc lattice with a 4 site basis.  The compactification is associated with the larger bcc lattice.  There are thus 4 states, so we have a 4 state generalization of the Ising model with $Z_4$ symmetry, which is known as the Ashkin Teller model.  This model has a parameter, which for different values gives, for example, the 4 state Potts model and the 4 state clock model.  It is not immediately obvious what the value of that parameter should be based on the form of \eqref{hthetaphi zk}.  However, from the analysis of the previous section, we know that the critical point is described by the $Z_4$ parafermion conformal field theory.  We thus expect that this model describes the Fateev Zamolodchikov point of the Ashkin Teller model.

For general $k$ the minima occur on a $k-1$ dimensional lattice formed by combinations of $\sqrt{2} \pi \vec d_a$.
There are $k$ distinct but equivalent minima to this potential, which can be located at
$\vec\theta_\sigma=\vec\theta_n = n \sqrt{2}\pi \vec d_1$, for $n=0, ..., k-1$.  The minima at $n=k$ is equivalent to the one at $n=0$ because from \eqref{d sum} $k \vec d_1 = \sum_{a=1}^k \vec d_{aj}$.  All other minima of $\cos \sqrt{2} \vec d_{ab}\cdot\vec\theta_\sigma$ can also be reduced to these $k$ minima with a suitable combination of $\sqrt{2}\pi \vec d_{ab}$.  This model thus describes a $k$ state system with $Z_k$ symmetry.  $Z_k$ models have extra parameters for $k \ge 4$, but as discussed above, since the critical point is described by the $Z_k$ parafermion theory we conclude that our model describes the Fateev Zamolodchikov point of the $Z_k$ model.

Now we consider the operators  $\Sigma_a$ discussed above, which has a simple interpretation.  When $(u_{ab}+v_{ab}\Upsilon^R_{ab}\Upsilon^L_{ab}$ opens a large gap, then we can restrict $\vec\theta_\sigma$ to the minima $\vec\theta_n = n \sqrt{2} \pi \vec d_1$.  It is then straightforward to see that
\begin{equation}
\Sigma^R_{j,a} \Sigma^L_{j,a} = e^{i\sqrt{2} \vec d_a\cdot\vec\theta_n} = e^{-2\pi i n/k},
\end{equation}
independent of $a$.  This is precisely the spin order parameter of the $Z_k$ model, which gives different values $e^{-2\pi i n/k}$ for the different ordered states specified by $n$.  We thus conclude that when the $[SU(2)_1]^k/SU(2)_k$ sector is gapped, the operator $\Sigma^R_{j,a}$, when projected into the $SU(2)_k/U(1)$ sector corresponds precisely to the spin field $\sigma$ of the $Z_k$ model.

\subsection{Generalization}

\begin{figure}
\includegraphics[width=3.3in]{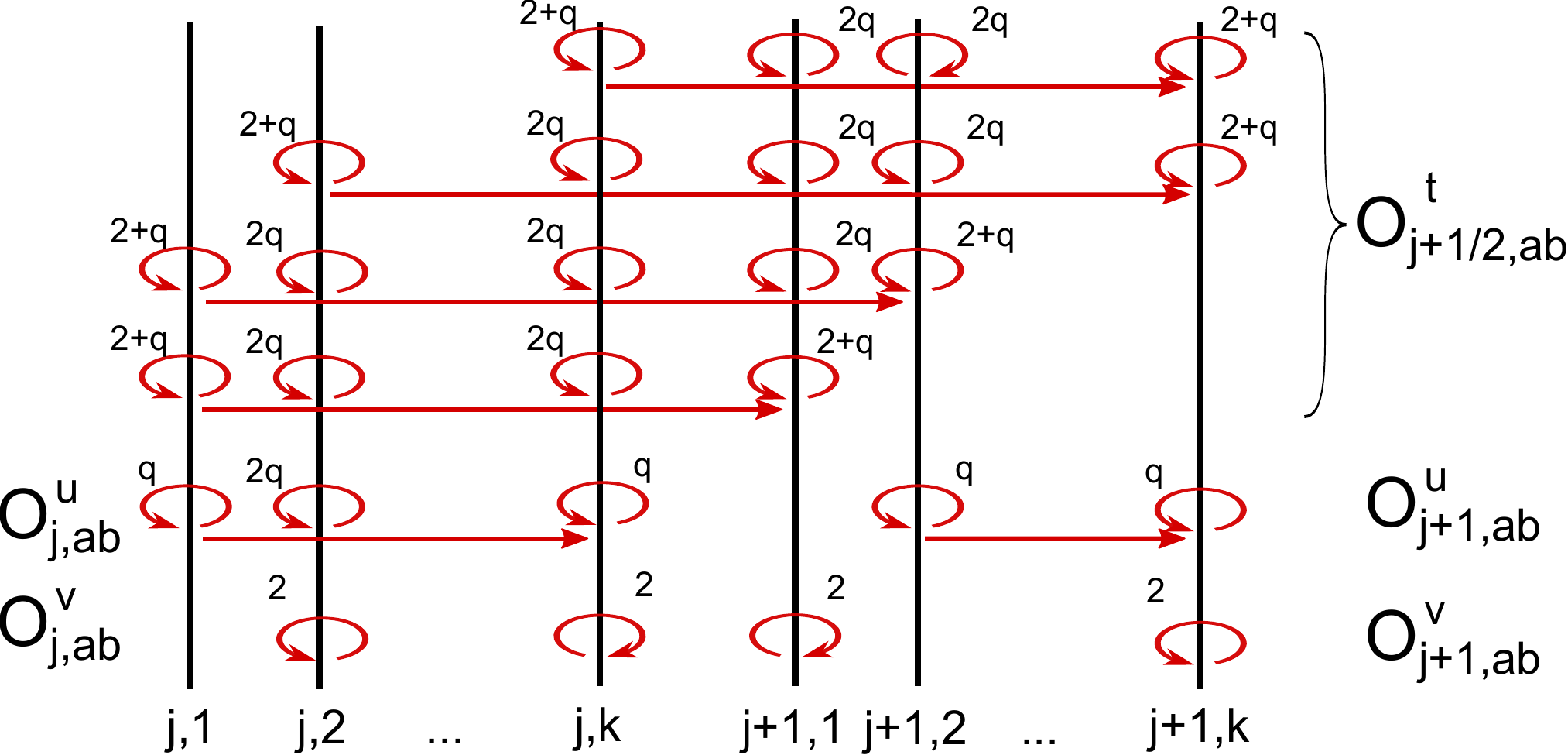}
\caption{Schematic diagram for the generalized Read Rezayi state at filling $\nu = k/(2 +k q)$.
groups of $k$ wires are coupled to one another by $O_{j+1/2,ab}^t$, which require a specific
staggered magnetic field for momentum conservation.  Wires within a group
are coupled by $O^{u,v}_{j,ab}$, which are independent of the field.  Representative
examples of $O_{j+1/2,ab}^t$ and $O^{u,v}_{j,ab}$ are shown.}
\label{Fig 10}
\end{figure}

As in Section III.C our construction for the Read Rezayi sequence can be generalized by introducing a staggered component to the field.  Again, the way to think about it is to start with the bosonic state at $\nu = k/2$, which can be viewed as a staggered field, in which the field between neighboring wires is $b$ for one out of every $k$ neighbors and $0$ for the other $k-1$ neighbors.  Keeping this staggered field fixed, we now add a uniform field $\bar b$, and find that for certain values, which correspond to filling factor
\begin{equation}
\nu = {k\over{2+ k q}}
\end{equation}
there are allowed tunneling processes, which have a structure similar to (\ref{v rr}-\ref{Ov rr}).  Expressed in terms of charge/neutral variables, as in \eqref{phirl rhosigma rr}, we find that the neutral sector is independent of $q$, while the charge sector is modified, as in \eqref{phirho mr commutator}.

Rather than repeating the algebra in Section III.C, we simply display the diagram, analogous to Fig \ref{Fig 7}, in Fig \ref{Fig 10}.

\section{Conclusion}
\label{sec:conclusion}

In this paper we have introduced a new formulation of non-Abelian fractional quantum Hall states, in which electronic models built from coupled interacting one dimensional wires can be analyzed using Abelian bosonization.  The picture that emerges from this analysis is summarized in Figs. \ref{Fig 6} and \ref{Fig 8}.  Non-Abelian states can be viewed as systems in which the original one dimensional chiral fermion modes are split into fractionalized sectors, in accordance with the coset construction of conformal field theory.  The different coset sectors are then coupled to one another in ``opposite directions".  This leads to a simple understanding of the edge states and quasiparticles that is similar in some respects to the one dimensional AKLT model.  In the case of the Moore Read state, the $c=1/2$ coset theory has a simple fermion representation.  For the more general Read Rezayi states, the coset theory can be identified by a mapping to the critical point of a $Z_k$ statistical mechanics problem.  We now conclude with some future directions and open problems.

It is natural to speculate that a coupled wire construction is possible for {\it any} quantum Hall state.
For example, we expect that it should be possible to construct the hierarchical generalizations of the Moore Read and Read Rezayi states\cite{bonderson}.  In addition, it would be interesting to develop the construction for the non-Abelian spin singlet state introduced by Ardonne and Schoutens\cite{ardonne}, which is based on a $SU(3)_2$ coset theory.  More generally, a much wider variety of quantum Hall states can be formulated using the parton construction.  It would be interesting to understand the connection between our more concrete approach - which can be formulated in terms of electrons and solved - and the more abstract constructions.    Our approach should also be applicable to spin models, and it is likely an anisotropic version of Kitaev's honeycomb lattice model\cite{kitaevhoneycomb} could be analyzed.

It would be interesting to further explore the way in which the coupled wire construction accounts for the non-Abelian statistics of the quasiparticles.  In the Abelian case, as explained in Section II.C.2, the Abelian statistics of the Laughlin quasiparticles can be simply understood.  It should be possible to understand the degeneracy and braiding properties associated with quasiparticles in the non-Abelian case in terms of the coupled wire model.

Finally, we ask whether the coupled wire construction could be useful for describing any phases which are not already well understood.  It seems likely that the coupled wire model could give a concrete representation of fractional Chern insulators, non-Abelian Chern insulators and fractional topological insulators in two, and possibly three dimensions.

\acknowledgments

We thank Maissam Barkeshli and Andreas Ludwig for helpful discussions.  We especially thank
Eduardo Fradkin for discussions and for introducing us to Ref. \onlinecite{fradkin}.
This work was supported by NSF grant DMR-0906175.

\begin{appendix}

\section{Klein Factors and Zero Modes}

In this appendix we carefully treat the Klein factors for fermions, along with the zero momentum modes of $\varphi$ and $\theta$.  This shows that in expressions such as \eqref{h0+vcos} and \eqref{h0sll+vcos h}, the Klein factors may be ignored.

An explicit formulation of the Klein factors can be formulated by considering a system with periodic boundary conditions in the $x$ direction.  The chiral modes may then be treated separately.  Define
\begin{equation}%\label{eq:AppKlein}
\psi^p_j(x) = {\kappa^p_j\over\sqrt{2\pi x_c}} e^{i\phi^p_j(x)}.
\label{psip}
\end{equation}
Here, $p = R/L = +1/-1$ describes the right and left moving chiral modes, and
\begin{equation}
[\phi^p_j(x),\phi^{p'}_{j'}(x')] = i p \pi\delta_{pp'}\delta_{jj'}s_{xx'},
\label{phi commutator s}
\end{equation}
where $s_{xx'}={\rm sgn}(x-x')$.  The Klein factors are necessary to assure that fermion fields associated with different channels anticommute.  They may be represented by the factors
\begin{equation}
\kappa^p_i=(-1)^{\sum_{(j,q)<(i,p)} N^q_j},
\label{kappap}
\end{equation}
where the number operator for each chiral channel
\begin{equation}
N^p_i = p \int dx \partial_x\phi^p_i/2\pi,
\end{equation}
satisfies $[N^p_j,\phi^{p'}_{j'}]=i\delta_{jj'}\delta_{pp'}$ and has integer eigenvalues.
We have chosen an ordering of the chiral modes with direction $p=R/L$ on wire $i$, such that
$(i,L) < (i,R) < (i+1,L) < (i+1,R)$.
Defined in this way, the Klein factors mutually commute $[\kappa^p_j,\kappa^{p'}_{j'}]=0$,
but the fermion operators anticommute, $\{\psi^p_j,\psi^{p'}_{j'}\}=0$.   Other choices for
the phase factors in \eqref{kappap} are also possible.
%Different choices correspond to different phase conventions for
%the zero mode states $|\{N^q_j\}\rangle$.  They are related by a unitary transformation that %modifies those phases.

The density and phase fields defined on each wire may be introduced as
\begin{eqnarray}
\varphi_j &=& (\phi^R_j + \phi^L_j + \pi N^L_j)/2\\
\theta_j &=& (\phi^R_j - \phi^L_j + \pi N^L_j)/2
\label{varphitheta n}
\end{eqnarray}
These satisfy $[\theta_j(x),\theta_{j'}(x')]=[\varphi_j(x),\varphi_{j'}(x')]=0$,
along with
\begin{equation}
[\theta_j(x),\varphi_{j'}(x')] = i\pi \delta_{jj'} \Theta(x-x').
\end{equation}
The electron operators are then
\begin{equation}
\psi^p_j(x) = {\kappa_j\over\sqrt{2\pi x_c}} e^{i(\varphi_j + p\theta_j)}.
\end{equation}
where the Klein factor (with no superscript) used in \eqref{bosonize psirl}, is
\begin{equation}
\kappa_i = (-1)^{\sum_{j<i} N_j^R + N_j^L}
\end{equation}
is now independent of $p$.

Consider the backscattering operator on an individual wire,
\begin{equation}
{\cal O}_j = \psi^{L\dagger}_j \psi^R_j = {{1\over{2\pi i x_c}} e^{2 i \theta_j}}
\end{equation}
The Klein factors are absent, and can safely be ignored.

In the subsections 1 and 2 we will apply this analysis to the Laughlin states and hierarchy states.  The bosonic Moore Read state does not require Klein factors in the original model, however, they require care when refermionizing.  This is discussed in subsection 3.

\subsection{Laughlin States}

The electron operator responsible for the fermionic Laughlin state $\nu=1/m$, for $m$ odd, is
\begin{equation}\label{eq:AppLaughlinelec}
\mathcal{O}^{1/m}_{\ell=j+1/2} =
(\psi_{j+1}^{L\dagger})^{m+1\over 2} (\psi_{j+1}^{R})^{m-1\over 2}
(\psi_j^{L\dagger})^{m-1\over 2} (\psi_j^{R})^{m+1\over 2}
\end{equation}
We write this as
\begin{equation}
{\cal O}^{1/m}_{j+1/2} = \tilde\psi_{j+1}^{L\dagger} \tilde\psi_{j}^{R}
\end{equation}
where
\begin{eqnarray}
\tilde\psi_{j}^R &=& (\psi^{L\dagger}_j)^{m-1\over 2}(\psi^R_j)^{m+1\over 2} \nonumber\\
\tilde\psi_{j}^L &=& (\psi^{R\dagger}_j)^{m-1\over 2}(\psi^L_j)^{m+1\over 2} .
\label{psitilde}
\end{eqnarray}

We now keep the Klein factors and define $\tilde\theta_{j+1/2}$ and $\tilde\varphi_{j+1/2}$ such that
\begin{eqnarray}
e^{2i \tilde\theta_{j+1/2}} &=& \tilde\psi_{j+1}^{L\dagger } \tilde\psi_{j}^{R} \nonumber\\
e^{2i \tilde\varphi_{j+1/2}} &=& \tilde\psi_{j+1}^{L} \tilde\psi_{j}^{R}.
\label{e^2itheta}
\end{eqnarray}
Then,
\begin{eqnarray}
2\tilde\theta_{j+1/2} &=& \tilde\phi^R_{j}-\tilde\phi^L_{j+1}   + \pi \tilde N^\theta_j \nonumber \\
2\tilde\varphi_{j+1/2} &=& \tilde\phi^R_{j}+\tilde\phi^L_{j+1} + \pi \tilde N^\varphi_j.
\label{thetavarphiN}
\end{eqnarray}
where
\begin{eqnarray}
\tilde\phi^R_{j} &=& {1+m\over 2}\phi^R_j + {1-m\over 2} \phi^L_j  \nonumber\\
\tilde\phi^L_{j} &=& {1-m\over 2}\phi^R_j + {1+m\over 2} \phi^L_j .
\end{eqnarray}
$\tilde N^{\theta/\varphi}_j$ are sums of $N_i^{R/L}$ determined by the Klein factors, using (\ref{psip},\ref{kappap},\ref{psitilde}).  Defined in this way, the commutation relations obeyed by $\tilde\psi_j^{R/L}$ guarantee that
$[e^{2i A_\ell(x)},e^{2i B_{\ell'}(x')}]=0$ for $A,B = \tilde\theta,\tilde\varphi$ and $x\ne x'$.  This means that $[A_\ell(x),B_{\ell'}(x')]= i P^{AB}_{\ell\ell'} \pi/2$, where $P^{AB}_{\ell\ell'}$ is an integer.
However, there is freedom in how $\tilde N^{\theta/\varphi}_j$ is defined because \eqref{kappap} is unchanged when $\tilde N^{\theta/\varphi}_j$ is increased by an even integer (which could depend on $N^{R/L}_i$).  This freedom can be exploited to define $\tilde\theta$ and $\tilde\phi$ so that they obey a standard commutation relation.  While a general method for determining $\tilde N^{\theta/\varphi}_j$ remains to be developed, we have by trial and error found (non unique) solutions.

For
\begin{eqnarray}
\tilde N^\theta_j &=& {m-1\over 2} N_j^L + N_j^R + {m-1\over 2} N_{j+1}^L \\
\tilde N^\varphi_j &=& {m-1\over 2} N_j^L + m N_j^R + {m-1\over 2} N_{j+1}^L,
\end{eqnarray}
the fields $\tilde\theta_j$ and $\tilde\varphi_j$ defined in \eqref{thetavarphiN} satisfy \eqref{psip},\eqref{kappap}, \eqref{psitilde} and \eqref{e^2itheta}, as well as
$[\tilde\theta_i(x),\tilde\theta_j(x')] = [\tilde\varphi_i(x),\tilde\varphi_j(x')] = 0$ and
\begin{equation}
[\tilde\theta_i(x),\tilde\varphi_j(x')] = i\pi\delta_{ij} \Theta(x-x').
\end{equation}

\subsection{Hierarchy States}

The procedure for defining the Klein factors for the hierarchy states is similar to that in the preceding section.  Here we just sketch the process.  We may again write the tunneling operators, defined in \eqref{Ohierarchy} as
\begin{eqnarray}
{\cal O}_{2k} = \tilde\psi_{k+1,1}^{L\dagger} \tilde\psi_{k,1}^{R} \nonumber\\
{\cal O}_{2k+1} = \tilde\psi_{k+1,2}^{L\dagger} \tilde\psi_{k,2}^{R}
\end{eqnarray}
with
\begin{eqnarray}
\tilde\psi_{k,1}^R  &=& (\psi^{R}_{2k-1})^{m_1+n\over 2} (\psi^{L\dagger}_{2k-1})^{m_1-n\over 2}
(\psi^{L\dagger}_{2k}\psi^{R}_{2k})^{m_0} \nonumber\\
\tilde\psi_{k,1}^L  &=& (\psi^{L}_{2k-1})^{m_1+n\over 2} (\psi^{R\dagger}_{2k-1})^{m_1-n\over 2} \\
\tilde\psi_{k,2}^R  &=& (\psi^{R}_{2k})^{m_1+n\over 2} (\psi^{L\dagger}_{2k})^{m_1-n\over 2} \nonumber\\
\tilde\psi_{k,2}^L  &=& (\psi^{L\dagger}_{2k})^{m_1+n\over 2} (\psi^{R\dagger}_{2k})^{m_1-n\over 2}(\psi^{R\dagger}_{2k-1}\psi^{L}_{2k-1})^{m_0} \nonumber
\end{eqnarray}
We now define
\begin{eqnarray}
e^{2i \tilde\theta_{k+1/2,a}} &=& \tilde\psi_{k+1,a}^{L\dagger} \tilde\psi_{k,a}^{R}  \nonumber\\
e^{2i \tilde\varphi_{k+1/2,a}} &=& \tilde\psi_{k+1,a}^{L} \tilde\psi_{a,k}^{R}.
\end{eqnarray}
Then,
\begin{eqnarray}
2\tilde\theta_{k+1/2,a} &=& \tilde\phi^R_{a,k+1}-\tilde\phi^L_{a,k}   + \pi \tilde N^\theta_{a,k}\nonumber \\
2\tilde\varphi_{k+1/2,a} &=& \tilde\phi^R_{a,k+1}+\tilde\phi^L_{a,k} + \pi \tilde N^\varphi_{a,k},
\end{eqnarray}
with
\begin{eqnarray}
\tilde\phi^R_{k,1} &=& {n+m_1\over 2} \phi^R_{2k-1} + {n-m_1\over 2} \phi^L_{2k-1}+ m_0 (\phi^R_{2k}-\phi^L_{2k}) \nonumber\\
\tilde\phi^L_{k,1} &=& {n-m_1\over 2} \phi^R_{2k-1} + {n+m_1\over 2} \phi^L_{2k-1}  \nonumber\\
\tilde\phi^R_{k,2} &=& {n+m_1\over 2} \phi^R_{2k} + {n-m_1\over 2} \phi^L_{2k}\\
\tilde\phi^L_{k,2}&=& {n-m_1\over 2} \phi^R_{2k} + {n+m_1\over 2} \phi^L_{2k} - m_0 (\phi^R_{2k-1}-\phi^L_{2k-1}). \nonumber
\end{eqnarray}
and
\begin{eqnarray}
\tilde N^\theta_{k+1/2,1} &=& {n-m_1\over 2}(N^L_{2k-1}+N^L_{2k+1}) - \nonumber \\ && (n+m_0) N^L_{2k}  -n(N^R_{2k-1}+N^R_{2k}) \nonumber \\
\tilde N^\theta_{k+1/2,2} &=& {n-m_1\over 2}(N^L_{2k}+N^L_{2k+2}) - \nonumber \\ && (n+m_0) N^L_{2k+1}  -n(N^R_{2k}+N^R_{2k+1}) \nonumber \\
\tilde N^\varphi_{k+1/2,1} &=& {n-m_1\over 2}(N^L_{2k-1}-N^L_{2k+1}) - \nonumber \\ && (n+m_0) N^L_{2k} -n(N^R_{2k-1}-N^R_{2k}) \\
\tilde N^\varphi_{k+1/2,2} &=& {n-m_1\over 2}(N^L_{2k}-N^L_{2k+2}) + \nonumber \\ && (n+m_0) N^L_{2k} -n(N^R_{2k}-N^R_{2k+1}).\nonumber
\end{eqnarray}

These fields then satisfy
\begin{equation}
[\tilde\theta_{k,a}(x),\tilde\theta_{l,b}(x')] = [\tilde\varphi_{k,a}(x),\tilde\varphi_{l,b}(x')] = 0.
\end{equation}
In addition
\begin{equation}
[\tilde\theta_{k,a}{x},\tilde\varphi_{l,b}(x')] = i\pi \delta_{kl} (K_{ab} \Theta(x-x') + W_{ab} ),
\end{equation}
with $K$ given in \eqref{k matrix} and $W_{ab}$ are integers.  It may be possible to find another choice for $\tilde N^{R/L}_{k,a}$ in which $W_{ab}=0$.  However, for the purpose of this paper, this choice suffices.

\subsection{Moore-Read state}

Here we provide the details of the zero modes and Klein factors for the $\nu=1/2$ bosonic Moore Read state discussed in Section III.A.   Since it is a model of bosons, there are no Klein factors in the original model.  However, it is necessary to keep track of the zero modes in order to correctly fermionize the theory.  The analysis in this appendix leads to the appropriate (and slightly counterintuitive) factors of $i$ in Eqs. (\ref{Ot+hc}-\ref{Ov+hc}).

Recall $\Phi^\dagger_{j,a}(x)\sim e^{i\varphi_{j,a}(x)}$ is the boson creation operator and $\rho_n\sim e^{in(2k_fx+2\theta(x))}$ represents the density wave at $q\sim2\pi\bar\rho n$.
In order for these fields to commute for $x\ne x'$, we require
\begin{equation}
[\theta_{j,a}(x),\varphi_{j',b}(x')] = i \pi \delta_{jj'}\delta_{ab}\Theta(x-x')
\end{equation}
where $\Theta(x-x')$ is the step function.  To define the chiral modes, with the appropriate commutation relations, we must augment \eqref{chiral mr} with the appropriate factors as in \eqref{varphitheta n}.  This may be written
\begin{eqnarray}
\theta_{j,a} &=& (\tilde\phi^R_{j,a} - \tilde\phi^L_{j,a} + \pi N^L_{j,a})/2\nonumber\\
\varphi_{j,a} &=& (\tilde\phi^R_{j,a} + \tilde\phi^L_{j,a} + \pi N^L_{j,a})/2.
\end{eqnarray}
Here, $N^p_{j,a} = p \int dx \partial_x\tilde\phi^p_{j,a}/(2\pi)$.  These fields obey commutation relations identical to \eqref{phi commutator s}.  The extra $N^L_{j,a}$ term was
necessary to make $\phi^R_{j,a}$ and $\phi^L_{j,a}$ commute.  Charge and spin modes can then be defined, as in \eqref{chiral mr rhosigma}, and the correction terms involve
$N^p_{j,\mu} = p \int dx \partial_x\tilde\phi^p_{j,\mu}/(2\pi)$, which satisfy
\begin{equation}
[N^p_{j,\mu},\tilde\phi^p_{j,\mu}] = i
\end{equation}
for $\mu = \rho,\sigma$.

In terms of these variables we then write
\begin{eqnarray}
{\cal O}^u_j &=& \exp[i (\tilde\phi^R_{j,\sigma}+\tilde\phi^L_{j,\sigma} + \pi N^L_{j,\sigma} )] \nonumber\\
{\cal O}^v_j &=& \exp[i (\tilde\phi^R_{j,\sigma}-\tilde\phi^L_{j,\sigma} + \pi N^L_{j,\sigma} )].
\label{Ouv with n}
\end{eqnarray}
We now separate the chiral modes in the exponential, keeping track of the commutator, and obtain
\begin{eqnarray}
{\cal O}^u_j  &=& i e^{i(\tilde\phi^R_{j,\sigma}+\pi N^L_{j,\sigma})} e^{ i\tilde\phi^L_{j,\sigma}} \nonumber\\
{\cal O}^v_j &=&  i e^{i(\tilde\phi^R_{j,\sigma}+\pi N^L_{j,\sigma})} e^{-i\tilde\phi^L_{j,\sigma}}.
\end{eqnarray}
It then follows that
\begin{eqnarray}
{\cal O}^u_j+ h.c. &=& i [\cos (\tilde\phi^R_{j,\sigma}+\pi N^L_{j,\sigma}) \cos\tilde\phi^L_{j,\sigma} \nonumber\\
&& \quad - \sin (\tilde\phi^R_{j,\sigma}+\pi N^L_{j,\sigma}) \sin\tilde\phi^L_{j,\sigma}] \nonumber \\
{\cal O}^v_j+ h.c. &=&  i [\cos(\tilde\phi^R_{j,\sigma}+\pi N^L_{j,\sigma}) \cos \tilde\phi^L_{j,\sigma} \\
&& \quad + \sin(\tilde\phi^R_{j,\sigma}+\pi N^L_{j,\sigma}) \sin \tilde\phi^L_{j,\sigma}] \nonumber
\end{eqnarray}
Thus, the Josephson and inter wire coupling terms are given in (\ref{Ou+hc},\ref{Ov+hc}).  A similar analysis gives (\ref{Ot+hc}).
The factor of $N^L_{j,\sigma}$ in \eqref{Ouv with n} provides precisely the necessary factor to define the Klein factor for the fermions in \eqref{fermionize}, as in Eqs. \ref{psip} and \ref{kappap}.

A similar analysis can be applied to the Read Rezayi states for general $k$.  For example, \eqref{ou d rr} is modified to be
\begin{eqnarray}
{\cal O}^u_{j,ab} &=& e^{i{\vec d}_{ab}\cdot(\vec\phi^R_{j,\sigma}+\vec\phi^L_{j,\sigma} + \pi \vec N^L_{j,\sigma})/\sqrt{2}} \nonumber\\
&=& i e^{i {\vec d}_{ab}\cdot (\vec\phi^R_{j,\sigma} + \vec N^L_{j,\sigma})} e^{i{\vec d}_{ab}\cdot {\vec \phi}^L_{j,\sigma}}.
\end{eqnarray}
where $\vec N^p_{j,\sigma} = p\int dx \partial_x{\vec\phi}^p_{j,\sigma}/2\pi$ and
$\vec d_{ab}=\vec d_a-\vec d_b$.

\section{Decoupling of Energy Momentum Tensor}

In this appendix we review how the energy momentum tensor for $[SU(2)_1]^k$ is decomposed into $[SU(2)_1]^k/SU(2)_k$, $SU(2)_k/U(1)$ and $U(1)$.

Since the notations in the conformal field theory literature and the condensed matter literature are somewhat different, we first review the translation between the two for a single non interacting (right moving) fermion mode.  In the CFT notation, this is written as
\begin{equation}
\psi(z) = :e^{i \phi(z)} :
\label{psiz cft}
\end{equation}
where $z$ is a complex coordinate and the colons denote normal ordering.  The chiral boson field $\phi$ satisfies
\begin{equation}
\langle \phi(z) \phi(w) \rangle = - \log (z-w)
\label{phicorrelator}
\end{equation}
so that $\psi$ satisfies $\langle \psi^\dagger(z) \psi(w)\rangle = 1/(z-w)$.
The energy momentum tensor is
\begin{equation}
T(z) = - {1\over 2} :(\partial_z\phi)^2:.
\label{t cft}
\end{equation}
Using \eqref{phicorrelator} and Wick's theorem, it can then be shown that the singular terms in the operator product expansion (OPE) of $T(z)$ with $\psi(w)$ are
\begin{equation}
T(z) \psi(w) = {\Delta\over {(z-w)^2}} \psi(w) + {1\over{z-w}}\partial_w \psi(w) .
\label{topepsi}
\end{equation}
with $\Delta=1/2$.  This shows that $\psi$ is a primary field with dimension $\Delta=1/2$.

In the condensed matter literature the chiral fermion field is often written as
\begin{equation}
\psi(x,\tau) = {1\over{\sqrt{2\pi x_c}}} e^{i\phi(x,\tau)},
\label{psixtau cmt}
\end{equation}
where the operator is not normal ordered and $x_c$ is a convergence factor in divergent momentum integrals, which plays the role of a short distance cutoff.  Since $\exp i\phi = (x_c/L)^{1/2}:\exp i\phi :$, the cutoff
$x_c$ can be eliminated by writing \eqref{psixtau cmt} using a normal ordered exponential.
$\psi$ satisfies $\langle \psi(x,\tau) \psi(0,0)\rangle = [2\pi(v_F\tau+i x)]^{-1}$.  The dynamics is governed by the Hamiltonian
\begin{equation}
H = \int dx {v_F\over{4\pi}} :(\partial_x\phi)^2:.
\label{h cmt}
\end{equation}

To make contact with the CFT notation, we first observe that
the normalization of $\psi$ in \eqref{psixtau cmt} differs by a factor of $\sqrt{2\pi}$ from \eqref{psiz cft}.  Consider a finite system with periodic boundary conditions of length $L$, so that $v_F\tau+ix$ is defined on a cylinder, and introduce the radial variable
\begin{equation}
z(x,\tau) = e^{2\pi(v_F\tau+ix)/L}.
\label{zxtau}
\end{equation}
The fermion field on the cylinder then has the form similar to \eqref{psiz cft},
\begin{equation}
\psi(x,\tau) = \sqrt{2\pi z\over L} : e^{i\phi(z)} :  |_{z = z(x,\tau)}.
\end{equation}
Aside from the $\sqrt{2\pi}$ difference in the normalization, this is equivalent to \eqref{psixtau cmt} for $L\rightarrow\infty$.

The Hamiltonian \eqref{h cmt} corresponds to the lowest mode of the energy momentum tensor in the radial quantization.
\begin{equation}
H = {2\pi v_F\over L} L_0,
\label{hl0}
\end{equation}
with
\begin{equation}
L_0 = {1\over{2\pi i}}\oint dz z T(z)
\label{l0t}
\end{equation}
where the integral is on a circle $|z|= e^{2\pi v_F \tau/L}$.  It can readily be seen that
\eqref{zxtau}, \eqref{hl0} and \eqref{l0t} imply that \eqref{t cft} and \eqref{h cmt} are equivalent.

Now consider the $k$ fields $\tilde\phi_a$, along with $\tilde\phi_{\mu = \rho,\sigma}$ considered in Section IV.A, which are related by \eqref{phirl rhosigma rr}.   (Again, we consider only a single (ie right moving) chiral sector, and omit the superscript $R$).   In the notation defined above, these satisfy
\begin{eqnarray}
\langle \tilde\phi_a(z) \tilde\phi_b(w) \rangle &=& - \delta_{ab} \log(z-w) \nonumber \\
\langle \tilde\phi_\mu(z) \tilde\phi_{\mu'}(w) \rangle &=& - \delta_{\mu\mu'} \log(z-w),
\end{eqnarray}
and the energy momentum tensor is
\begin{eqnarray}
T_{[SU(2)_1]^k} &=& - {1\over 2} \sum_a :(\partial_z\tilde\phi_a)^2 :  \\
&=& - {1\over 2}:\left( (\partial_z\tilde\phi_\rho)^2 + (\partial_z\vec\phi_\sigma)^2 \right):
\label{tphirho}
\end{eqnarray}
For each of the $k$ channels, the operators
\begin{eqnarray}
J_a^z &=& i \partial_z\tilde\phi_a  \nonumber \\
J_a^\pm &=& J_a^x \pm i J_a^y = \sqrt{2} e^{\pm i \sqrt{2}\tilde\phi_a}
\label{jabelian}
\end{eqnarray}
form an $SU(2)_1$ current algebra.
% characterized by the OPE
%\begin{equation}
%J_a^i(z) J_b^j(w) = \delta_{ab}\left( {i \epsilon^{ijk}\over{z-w}} J_a^k(w) + %{\delta_{ij}/2\over{(z-w)^2}}\right).
%\label{kmalgegra}
%\end{equation}
Using the fact that $:{\bf J}_a\cdot {\bf J}_a: = -3:(\partial_z\phi_a)^2:$, we may write the energy momentum tensor in a way that reflects the $[SU(2)_1]^k$ symmetry,
\begin{equation}
T_{[SU(2)_1]^k} = {1\over 6} \sum_a :{\bf J}_a\cdot {\bf J}_a:.
\end{equation}

The diagonal subalgebra generated by ${\bf J} = \sum_a {\bf J}_a$ forms a
$SU(2)_k$ current algebra.
%, in which \eqref{kmalgegra} is modified to be
%\begin{equation}
%J^i(z) J^j(w) = \delta_{ab}\left( {i \epsilon^{ijk}\over{z-w}} J^k(w) + {%\delta_{ij}k/2\over{(z-w)^2}}\right).
%\end{equation}
The corresponding energy momentum tensor will be proportional to
${\bf J} \cdot {\bf J}$  (we now suppress the normal ordering symbols, for brevity).  The coefficient can be deduced by using \eqref{jabelian} to compute
\begin{equation}
{\bf J} \cdot {\bf J} = -(k+2)(\partial_z\tilde\phi_\rho)^2 - 2(\partial_z\phi_\sigma)^2
 + 2\sum_{a\ne b} e^{i\sqrt{2}(\vec d_a-\vec d_b)\cdot \vec\phi_\sigma}
\label{jdotj}
\end{equation}
If we require that the $(\partial_z\phi_\rho)^2$ terms in \eqref{tphirho} and \eqref{jdotj} coincide, then
an expression similar to \eqref{topepsi} shows that $J_z = i\sqrt{2k}\partial_z \phi_\rho$ has the appropriate scaling dimension $\Delta=1$.  We then recover the Sugawara energy momentum tensor,
\begin{equation}
T_{SU(2)_k} = {1\over{2(k+2)}} : {\bf J} \cdot {\bf J} :.
\label{tsu2k}
\end{equation}

It follows that we may express $T_{[SU(2)_1]^k}$ in \eqref{tphirho} as $T_{SU(2)_k} + T_{[SU(2)_1]^k/SU(2)_k}$, where
$T_{SU(2)_k}$ is given in \eqref{tsu2k}, and $T_{[SU(2)_1]^k/SU(2)_k}$ is the remainder.  Moreover,
$T_{SU(2)_k}$ may be further decomposed into $T_{U(1)} + T_{SU(2)_k/U(1)}$, where the $U(1)$ term, generated by $J^z$ is simply the $(\partial_z\tilde\phi_\rho)^2$ term in \eqref{jdotj}, and $T_{SU(2)_k/U(1)}$ is the rest.  This leads to the final results quoted in (\ref{tu1},\ref{tparafermion},\ref{tcoset}).

\section{Operator Product Expansions}

In this appendix we provide the details of the calculations for Eqs. \ref{ope Psi1} - \ref{ope Upsilon 2} that show that the operators
$\Psi$ and $\Upsilon$ defined at the end of Section IV.A are primary fields of the $SU(2)_k/U(1)$
$Z_k$ parafermion sector and the $[SU(2)_1]^k/SU(2)_k$ sectors respectively.  The key is to compute the singular terms in the operator product expansion of the energy momentum tensors for the coset sectors (given in Eqs. \ref{tu1}, \ref{tparafermion}, \ref{tcoset}  and discussed in the previous Appendix) with these operators.

The necessary terms involve two kinds of products, which it is useful to consider separately.  First, using Wick's theorem and \eqref{phicorrelator}, the OPE of the quadratic terms in $T$ with an exponential operator takes the form
\begin{eqnarray}
-{1\over 2}&(\partial_z\vec\phi_\sigma(z))^2 e^{i \vec D \cdot \vec \phi_\sigma(w)} \hspace{1in} \nonumber \\
&=  \left( { |\vec D|^2/2 \over {(z-w)^2}}
+ {i \vec D \cdot \partial_z \vec\phi_\sigma(z)\over {z-w}}\right) e^{i \vec D \cdot \vec \phi_\sigma(w)} \label{ope1}\\
&= \left( { |\vec D|^2/2 \over {(z-w)^2}}
+ {\partial_w \over {z-w}}\right) e^{i \vec D \cdot \vec \phi_\sigma(w)} + ...
\nonumber
\end{eqnarray}
For brevity we have suppressed the normal ordering symbols.  This shows that the operator $e^{i \vec D \cdot \vec \phi_\sigma}$ is a primary field of the $[SU(2)_1]^k$ theory (described by \eqref{tphirho}) with scaling dimension $\Delta = |\vec D|^2/2$.  In particular, using \eqref{d prod}, this shows that $\Delta_\Psi = |\vec d_a|^2 = 1 - 1/k$ and
$\Delta_\Upsilon = \Delta_\Xi = |\vec d_a - \vec d_b|^2/4 = 1/2$.

Using the Baker Hausdorf formula and \eqref{phicorrelator}, we may show that
\begin{eqnarray}
e^{i \vec D_1 \cdot \vec\phi_\sigma(z)}&e^{i \vec D_2 \cdot \vec\phi_\sigma(w)} \hspace{2.2in} \nonumber\\
=&{1\over {(z-w)^{-\vec D_1 \cdot \vec D_2}}} e^{i (\vec D_1 \cdot\vec\phi_\sigma(z) + \vec D_2 \cdot\vec\phi_\sigma(w))} \hspace{.4in}\label{ope2}\\
=&{1 + i(z-w) \vec D_1 \cdot \partial_w \vec\phi_\sigma(w)\over {(z-w)^{-\vec D_1 \cdot \vec D_2}}}  e^{i (\vec D_1 + \vec D_2)\cdot\vec\phi_\sigma(w)} + ...
\nonumber\end{eqnarray}

The OPE's for $\Psi$ involve \eqref{ope1} with $\vec D = \sqrt{2} \vec d_a$, along with \eqref{ope2} with
$\vec D_1 = \sqrt{2}(\vec d_a - \vec d_b)$ and $\vec D_2 = \sqrt{2} \vec d_c$.  In this case,
$\vec D_1 \cdot \vec D_2 = 2( \delta_{ac} - \delta_{bc})$, so there are singular terms in the OPE only for $c = b \ne a$.  Using the fact (from Eq. \ref{d sum}) that $\sum_{b\ne a} \vec d_a - \vec d_b = k \vec d_a$, we then find
\begin{eqnarray}
\sum_{a \ne b}& &e^{i\sqrt{2} (\vec d_a - \vec d_b)\cdot\vec\phi_\sigma(z)} \sum_c
e^{i\sqrt{2} \vec d_c \cdot \vec\phi_\sigma(w)}  \hspace{.5in}\nonumber \\
&=&\sum_{a \ne b}  \left( {1\over {(z-w)^2}} + {i\sqrt{2}(\vec d_a - \vec d_b)\cdot\partial_w\vec\phi_\sigma\over{z-w}} \right)
e^{i \sqrt{2}\vec d_a \cdot\vec\phi_\sigma} \nonumber\\
&=&\left( {k-1\over {(z-w)^2}} + {k \partial_w \over{z-w}} \right)
\sum_a e^{i \sqrt{2}\vec d_a \cdot\vec\phi_\sigma}.
\label{ope3}
\end{eqnarray}
Combining \eqref{ope1} (with $|\vec D|^2/2= 1-1/k$) and \eqref{ope3}, leads immediately to the results \eqref{ope Psi1} and \eqref{ope Psi2} quoted in Section IV.B.  That $T_{U(1)} \Psi = 0$ is obvious because $\Psi$ does not depend on $\tilde\phi_\rho$.

Thus, we have established that $\Psi$ is a primary field of the $SU(2)_k/U(1)$ sector.
$\Psi$ are a bosonized representation for $Z_k$ parafermion operators\cite{griffin}.  They can be combined to define more general operators,
\begin{equation}
\Psi_\ell = A_{k,l} \sum_{a_1 < ... < a_l} e^{i \sqrt{2}\sum_{i=1}^l \vec d_{a_i} \cdot\vec\phi_\sigma}
\end{equation}
For an appropriate normalization $A_{k,l} = \sqrt{l!(k-l)!/k!}$ these can be shown using \eqref{ope2} to satisfy the OPE's for $Z_k$ parafermions discovered by Zamolodchikov and Fateev\cite{zamolodchikov}.

The OPE's for $\Upsilon$ and $\Xi$ involve \eqref{ope1} with $\vec D = (\vec d_a-\vec d_b)/\sqrt{2}$ ($|\vec D|^2 = 1$), along with \eqref{ope2} with $\vec D_1 = \sqrt{2}(\vec d_a - \vec d_b)$ and $\vec D_2 = (\vec d_c-\vec d_d)/\sqrt{2}$.  It follows that $\vec D_1 \cdot \vec D_2 = \delta_{ac} + \delta_{bd} - \delta_{ad} - \delta_{bc}$.  This leads to a $1/(z-w)^2$ singularity for $a=d$ and $b=c$.   In addition, there is a
$1/(z-w)$ singularity for $a=d \ne b \ne c$ or $b=c \ne a \ne d$.
After an analysis similar to \eqref{ope3} it follows that
\begin{eqnarray}
\sum_{a \ne b, c \ne d} e^{i\sqrt{2}(\vec d_a - \vec d_b)\cdot \vec\phi_\sigma(z)}
e^{\pm i (\vec d_c - \vec d_d) \cdot\vec\phi_\sigma(w)/\sqrt{2}} \hspace{.4in}\nonumber \\
=  \left( {1\over{(z-w)^2}} + {2 \partial_w\over{z-w}}\right)
\sum_{a \ne b} e^{\mp i (\vec d_a - \vec d_b)\cdot \vec\phi_\sigma/\sqrt{2}} \label{ope4}\\
+ \sum_{c \ne a \ne b} {2\cos\left( (2 \vec d_c - \vec d_a-\vec d_b)\cdot \vec\phi_\sigma/\sqrt{2}\right) \over (z-w)} \nonumber
\end{eqnarray}
Since the last term is independent of the sign in the exponent, it cancels for $\Upsilon$.  Combining \eqref{ope4} with \eqref{ope1} (with $|\vec D|^2/2 = 1/2$ then leads to \eqref{ope Upsilon 1} and \eqref{ope Upsilon 2}.   Again, the $U(1)$ term vanishes because $\Upsilon$ is independent of $\tilde\phi_\rho$.  Thus, $\Upsilon$ is a primary field of the $[SU(2)_1]^k/SU(2)_k$ sector.

For $\Xi$, the last term does not cancel.  The OPEs for both $T_{SU(2)_k/U(1)}$ and $T_{[SU(2)_1]^k/SU(2)_k}$ are both non zero and do not have the form of a primary field.
Presumably, $\Xi$ can be written as a sum of terms that are products of primary fields in the
$SU(2)_k/U(1)$ and $[SU(2)_1]^k/SU(2)_k$ sectors.

\end{appendix}

\end{document}